\documentclass[11pt,letterpaper]{article} 
\pdfoutput=1
\usepackage{amsmath,amssymb,amsfonts}

\usepackage[usenames,dvipsnames]{xcolor}
\definecolor{airforceblue}{rgb}{0.36, 0.54, 0.66}
\definecolor{meb}{rgb}{0.01, 0.31, 0.59}
\usepackage[T1]{fontenc}
\def\title{The Phase Space of Gravity on Null Hypersurfaces} 
\usepackage{graphicx}
\usepackage[colorlinks=true, pdfstartview=FitV, linkcolor=blue, citecolor=magenta, urlcolor=blue]{hyperref}
\usepackage{cancel}
\usepackage{mdframed,framed}
\usepackage[usenames,dvipsnames]{xcolor}
\usepackage[normalem]{ulem}
\usepackage{layout}
\usepackage[nosort]{cite}

\usepackage{textgreek}

\newcommand{\beq}{\begin{eqnarray}}
\newcommand{\eeq}{\end{eqnarray}}
\newcommand{\beqn}{\begin{eqnarray}}
\newcommand{\eeqn}{\end{eqnarray}}
\newcommand{\bee}{\begin{equation} \begin{aligned}}
\newcommand{\eee}{ \end{aligned} \end{equation}}
\newcommand{\pa}{\partial}

\newcommand{\RR}{\mathbb{R}}

\newcommand{\cL}{{\cal L}}
\newcommand{\cO}{{\cal O}}

\newcommand{\fL}{\mathfrak{L}}

\newcommand{\olD}{{\overline{D}}}

\newcommand{\ttau}{\tilde{\tau}}

\newcommand{\nn}{\nonumber}

\usepackage{tikz-cd}
\usepackage{pict2e}
\makeatletter
\newcommand{\variable@rule}[1]{%
  \fontdimen8  
  \ifx#1\displaystyle\textfont3\else
    \ifx#1\textstyle\textfont3\else
      \ifx#1\scriptstyle\scriptfont3\else
        \scriptscriptfont3\relax
  \fi\fi\fi
}
\usepackage{mathtools}

\usepackage{tocloft}
\newcommand{\ve}{\varepsilon}

\newcommand{\cN}{{\cal{N}}}
\newcommand{\cC}{{\cal{C}}}
\newcommand{\cA}{{\cal{A}}}
\newcommand{\cF}{{\cal{F}}}

\newcommand{\paleft}{\stackrel{\rightarrow}{\pa}}

\newcommand{\bq}{{\overline{q}}}

\newcommand{\oD}{{\overline{D}}}

\newcommand{\rd}{\text{d}}

\newcommand{\ta}{{\tilde{\varepsilon}}}

\newcommand{\chkM}{{\color{red} \,\checkmark\kern-5pt{}_{M}}}

\newcommand{\be}{\begin{equation}}
\newcommand{\ee}{\end{equation}}
\newcommand{\bea}{\begin{eqnarray}}
\newcommand{\eea}{\end{eqnarray}}

\def\pa{\partial}

\newcommand{\pmu}{{{\mu}}}

\newcommand{\psigma}{{\tilde{\sigma}}}
\newcommand{\ppi}{{\tilde{\pi}}}

\newcommand{\del}[2]{\tilde\delta^{(3)}(x_{#1},x_{#2})}
\newcommand{\delc}[2]{\delta^{(3)}(x_{#1},x_{#2})}

\newcommand{\PB}[2]{\{ #1 , #2 \}}

\usepackage{ragged2e}
\usepackage{blindtext}

\newcommand{\perimeter}[1]{
	\centerline{
		\begin{minipage}[c]{0.7\textwidth}
			\begin{center}
			$^a$ Perimeter Institute for Theoretical Physics,\\
			 31 Caroline St. N., Waterloo ON, Canada, N2L 2Y5
			\end{center}
		\end{minipage}
		}
	}
\newcommand{\uiuc}[1]{
	\centerline{
		\begin{minipage}[c]{0.9\textwidth}
			\begin{center}
		$^b$ Illinois Center for Advanced Studies of the Universe \& Department of Physics,\\
			University of Illinois, 1110 West Green St., Urbana IL 61801, U.S.A.
			\end{center}
		\end{minipage}
		}
	}

\begin{document}

{\centering
 \vspace*{1cm}
\textbf{\LARGE{\title{}}}
\vspace{0.5cm}
\begin{center}
Luca Ciambelli,$^a$ Laurent Freidel,$^a$ Robert G. Leigh$^{b,a}$\\
\vspace{0.5cm}
\textit{\perimeter{}}\\
\vspace{0.5cm}
\textit{\uiuc{}}
\end{center}
\vspace{0.5cm}
{\small{\href{mailto:ciambelli.luca@gmail.com}{ciambelli.luca@gmail.com}, \ \href{mailto:lfreidel@perimeterinstitute.ca}{lfreidel@perimeterinstitute.ca}}}\\
\vspace{1cm}
\textit{This work was initiated and pursued together with our dear friend and colleague, Rob Leigh, whose recent passing has left us with a profound sense of loss. Rob’s insight, generosity of thought, and intellectual presence will be deeply missed.}\vspace{0.5cm}

\textit{Many of the ideas underlying this project took shape during a research focus group in Banff, where the three of us shared countless long and memorable discussions. Those conversations, filled with curiosity, insight, and friendship, remain an inseparable part of this work.}\vspace{0.5cm}

\textit{We dedicate this work to Rob’s memory, with gratitude for all that he gave us, both as a physicist and as a friend. We are certain that his influence will endure, through the ideas he helped shape, the people he inspired, and the lasting legacy he leaves in physics.}
\vspace{1cm}
\begin{abstract}
\vspace{0.5cm}
We construct the bulk kinematical Poisson structure of general relativity on a caustic-free null segment, prior to imposing the Raychaudhuri and Damour constraints. 
We first show that shifts of the Ehresmann connection, also known as Carroll boosts, are pure gauge in gravity. This allows the Ehresmann connection to be fixed as a background structure and leads to the prime phase space of null gravitational data.
 The resulting phase space comprises of a spin-0 pair that includes the area and its conjugate, a spin-1 pair including the null direction and its conjugate, and the spin-2 sector encoded in the  unimodular transverse metric. 
We then construct the brackets through  a three-stage Dirac reduction implementing a normalization, horizontality, and shear constraints. We find that the spin-2 bracket is non-local and involves an antisymmetric Green kernel for a first-order transport operator along the null generators.
The self-brackets of the spin-0 and spin-1 momenta involve, respectively, the vertical and horizontal derivatives of the transverse metric contracted with the spin-2 propagator.
We finally provide an independent derivation of the brackets from the explicit realization of the Hamiltonian vector fields of the symplectic form. These results provide the classical canonical arena for studying the null constraint algebra, observables, and quantization.
\end{abstract}}

\thispagestyle{empty}

\newpage
\tableofcontents
\thispagestyle{empty}
\newpage
\clearpage
\pagenumbering{arabic} 

\section{Introduction}

The canonical formulation of general relativity has played a central role in our understanding of gravitational physics \cite{Arnowitt:1962hi}. By identifying the phase space of the theory, its symplectic structure, its constraints, and the Poisson algebra of observables, it provides the natural classical starting point for any attempt at quantization \cite{Dirac:1958sc, dewitt1967quantum, Isham:1992ms}. This perspective is not merely formal: it clarifies which data are physical, which transformations are gauge, and how gravitational dynamics is encoded in constraint equations. In the standard spacelike formulation, this program has led to a detailed understanding of the ADM phase space and of the Hamiltonian and momentum constraints \cite{Arnowitt:1962hi, Misner:1973prb}. For null hypersurfaces, however, the analogous canonical picture is subtler and considerably less complete.

Null hypersurfaces occupy a privileged place in gravitational physics. They describe causal boundaries, horizons, light fronts, asymptotic radiation, and the characteristic surfaces along which gravitational information propagates. The gravitational data on a null hypersurface naturally separate into the area element, the shear, the expansion, and the momentum aspect, while the Einstein equations reduce to constraint equations intrinsic to the null surface, most notably the Raychaudhuri \cite{PhysRev.98.1123, Sachs:1961zz,Sachs:1962zzb} and Damour  \cite{Damour1979, damour1978black} equations, see also \cite{price1986membrane}.\footnote{The Carrollian nature of these equations has been appreciated in \cite{Donnay:2019jiz}, and studied in the fluid/gravity context in \cite{Redondo-Yuste:2022czg, Freidel:2022bai,Ciambelli:2023mvj}.} These equations govern the focusing of null generators and the transport of momentum along the hypersurface, and are therefore central to the classical physics of radiation, black holes, gravitational memory, and null boundaries \cite{Hawking:1973uf, Ashtekar:1981bq, Donnay:2015abr}.

Despite this importance, the canonical structure of gravity on null hypersurfaces remains less systematically understood than its spacelike counterpart. One reason is geometric \cite{Geroch:1977big,Torre:1985rw,Mars:1993mj,Gourgoulhon:2005ng}: the induced metric on a null hypersurface is degenerate, so the usual spacelike decomposition cannot be directly imported. Instead, the intrinsic geometry is Carrollian \cite{Henneaux:1979vn,Donnay:2019jiz,Ciambelli:2019lap}, see the review \cite{Ciambelli:2025unn} for more context and references. A null hypersurface carries a degenerate metric, a preferred null direction, and an auxiliary choice of horizontal structure. These ingredients come with nontrivial redundancies, including diffeomorphisms, rescalings of the null generator, and shifts of the Ehresmann connection. A proper canonical treatment must therefore distinguish genuine phase-space data from pure gauge structure before computing the Poisson brackets.

A second reason is conceptual. The Einstein equations on a null hypersurface are constraints, and one may ask whether the canonical brackets should be constructed after imposing these equations, or before doing so. In the pioneering works of Reisenberger \cite{Reisenberger:2007ku, Reisenberger:2012zq, Reisenberger:2018xkn}, the null constraint equations were incorporated in the construction of the free null initial-data phase space. These works also addressed the appearance of caustics and developed methods for defining Poisson brackets in their presence. In this work, following the direction taken in \cite{Ciambelli:2023mir}, we instead adopt a kinematical viewpoint.

The brackets derived here are classical brackets on the gravitational phase space of null data, prior to imposing the null constraint equations as dynamical restrictions. This strategy is motivated by two related considerations. First, the kinematical phase space is the arena on which the Raychaudhuri and Damour constraints act (see, e.g., \cite{Chandrasekaran:2018aop,Chandrasekaran:2023vzb, Odak:2023pga}); its Poisson structure must be known before the constraint algebra and the associated reductions can be analyzed. Second, although the present work is entirely classical, possible quantum applications strongly suggest that the dynamical equations should not be built into the phase space too early. At the quantum level, constraint algebras may acquire anomalous terms \cite{Ciambelli:2024swv, Freidel:2025ous} or require nontrivial operator orderings \cite{Freidel:2026stu}. A clean classical kinematical bracket is therefore the appropriate object to isolate first. Recent developments have also revealed a remarkable ultralocal structure of null dynamics, in which the theory organizes itself along individual null generators. Related null-algebraic structures play a central role in formulations and proofs of the generalized second law 
\cite{Wall:2011hj,Faulkner:2024gst}.

Previous analyses have clarified several aspects of the null phase space and its quantization, in particular the spin-zero sector associated with the area element and the expansion, and its relation to the Raychaudhuri equation \cite{Ciambelli:2023mir, Wieland:2024dop, Wieland:2025qgx, Ciambelli:2025flo,Ciambelli:2025fbo,Kowalski-Glikman:2026bwg}.\footnote{The analysis in \cite{Ciambelli:2026pwi} contains a spin-1 sector entirely dictated by the spin-0 breathing mode.} Algebraic approaches to gravitational observables associated with horizons and null subregions have explored various subsectors in \cite{Kudler-Flam:2023qfl, Klinger:2026tws}, while \cite{Ciambelli:2026vxa} addresses the quantization of the null characteristic initial-value problem. The canonical decomposition into spin-zero, spin-one, and spin-two pairs was identified in \cite{Hopfmuller:2016scf}, while its relation to null conservation laws and Hamiltonian generators was further developed in \cite{Hopfmuller:2018fni}. Complementary constructions of null-boundary phase spaces, their slicings, charges, and solution spaces were developed in \cite{Adami:2021nnf,Adami:2023wbe}. In particular, the symplectic reduction performed in \cite{Adami:2023wbe} aligns closely with the objectives of this paper.

These results provide complementary pieces of the null canonical picture and bring the problem close to the one considered here. The remaining step is to formulate a single gauge-independent kinematical phase space in which the spin-zero, spin-one, and spin-two sectors are simultaneously retained before the null Einstein constraints are imposed. This is the problem we address in the present work.

Concretely, this requires keeping track not only of the area and expansion, but also of the spin-two shear degrees of freedom and the spin-one momentum aspect governed by the Damour equation. It also requires treating the shift symmetry of the Ehresmann connection, also known as a Carroll boost, carefully. A key result of this work is that this symmetry is pure gauge in gravity: the Ehresmann connection can therefore be gauge-fixed at will, without freezing the physical diffeomorphisms of the null hypersurface. This leads to a prime phase space in which the symplectic structure is nondegenerate on the relevant gravitational data.

The purpose of this paper is to derive the complete kinematical Poisson structure of gravity on a null hypersurface, within the local null-segment setting considered here. Starting from the covariant symplectic potential, we perform the reduction to the prime phase space and identify the elementary canonical variables in the spin-zero, spin-one, and spin-two sectors. The resulting brackets exhibit the canonical role of the null momenta: while the spin-zero momentum involves  reparametrizations along the null generators of the metric, the spin-one momentum involves horizontal deformations, corresponding to spatial diffeomorphisms on the cuts. Together with the spin-two shear sector, these brackets provide the full classical bracket structure underlying the Raychaudhuri and Damour constraints. Moreover, we find that the nonlocality of the spin-two sector is encoded in a bilocal Green kernel transporting the metric data along the null generators. We obtain the brackets through a three-step Dirac reduction and independently reconstruct them from their Hamiltonian generators.

In this sense, the present work completes the local kinematical canonical analysis of the full null gravitational data within the setting considered here. It provides the framework in which the Raychaudhuri and Damour constraints, their reductions, the resulting observables, and possible quantizations can now be systematically studied. The dynamical questions begin only after this structure is in place: the brackets derived here are the classical arena in which the Raychaudhuri and Damour constraints, their reductions, and their observable consequences can be systematically studied.

Section \ref{s2} reviews the ruled Carrollian geometry and its intrinsic constraints. Section \ref{s3} establishes the prime phase space, while Section \ref{s4} derives the brackets through Dirac reduction. Section \ref{s5} reconstructs them from Hamiltonian generators, and Section \ref{s6} discusses the dynamical and quantum directions opened by these results.

\section{Null Hypersurfaces}\label{s2}

We review the main ingredients of gravitational physics on a null hypersurface. Following \cite{Ciambelli:2023mir} (see also the review \cite{Ciambelli:2025unn}), we start with the geometric notion of Carrollian and ruled Carrollian structures. We then discuss the choice of Carrollian connection, and the derivation of the Einstein constraints on the null hypersurface. We finish with an overview of the symmetries involved. While this discussion can all be done from a bulk ambient space viewpoint, we refrain from using the bulk as much as possible, in the holographic spirit of defining an intrinsic null dynamics for the geometric data. 

\subsection{Geometry}\label{S21}

We consider a $4$-dimensional bulk spacetime and a $3$-dimensional null hypersurface $\cN$. The latter is intrinsically described as a manifold endowed with a Carrollian structure $(q_{ab},\ell^b)$, where $\ell^a$ is a nowhere-vanishing vector field and $q_{ab}$ is a rank-$2$ metric such that $\ell^a q_{ab}=0$. The index $a$ is a $3$-dimensional index on $\cN$. The degenerate metric $q_{ab}$ decomposes into a unimodular part $\bq_{ab}$ and the infinitesimal area element $\Omega$, such that 
\beq\label{uni}
q_{ab}=\Omega \bq_{ab}.
\eeq

This definition of a Carrollian structure is unique up to rescaling $\ell \rightarrow \mathrm{e}^\lambda\ell$. This means that we have a conformal class of Carrollian vector fields $\ell^a$, a fact we exploited in \cite{Ciambelli:2024swv} to quantize the Raychaudhuri constraint.
Furthermore, generic null congruences may develop caustics (see the recent work \cite{Gadioux:2023pmw}),  where the light rays converge. We restrict attention to a null segment lying between caustics. The Carrollian structure thus describes the portion of a null hypersurface between caustics, where $\ell^a q_{ab}=0$ implies that $q_{ab}$ is degenerate. If we drop caustics from our analysis, we will correspondingly assume that $\cN$ is an open set with product topology $S\times I$, with $I$ an interval of the real line.

Given a Carrollian structure we can construct the expansion tensor, also known as extrinsic curvature, given by the Lie derivative along $\ell$ of the metric
\be 
\theta_{ab}= \frac12 \cL_{\ell} q_{ab}.
\ee 
We choose a Carrollian connection $D_a$ that is torsionless and minimally non-metrical. This means that ($D_\ell=\ell^a D_a$) 
 \be \label{Dq}
 D_\ell q_{ab}=0, \qquad \ell^c D_a q_{bc}  =-\theta_{ab}.
 \ee 
These properties guarantee that we can write the expansion tensor \`a la Brown-York (see \cite{Chandrasekaran:2021hxc}), 
\beq
\theta_{ab}=\frac12 \cL_\ell q_{ab}= D_{(a}\ell^c q_{b)c}.
\eeq
It is important to appreciate that when the expansion tensor $\theta_{ab}$ is non vanishing, it is not possible to have a Carrollian connection which is both torsionless and metrical, see \cite{Ciambelli:2025unn} and references therein. 
This follows from the key identity
\be 
\theta_{ab}  
= \tfrac12 N_{\ell ab} - N_{(ab) \ell} + T_{\ell (ab)}, 
\ee
where $N_{abc}= D_a q_{bc}$ is the non-metricity tensor and $T_{abc}= 2\Gamma_{[ab]}^d q_{dc}$ is the torsion.
This means that we must have either torsion or non-metricity, otherwise we constrain the expansion tensor. The case of torsion was investigated, e.g., in \cite{Figueroa-OFarrill:2020gpr}. The torsionless option introduced in \cite{Chandrasekaran:2021hxc,  Freidel:2022vjq, Ciambelli:2023mir}, and described here, relates the Carrollian connection to the pullback of the bulk Levi-Civita connection via the null rigging formalism \cite{Mars:1993mj}. Intrinsically, this is the standard Carrollian connection \cite{Ciambelli:2025unn}.

The Carrollian structure possesses on any cut $\cC$ of $\cN$ a  non-degenerate metric. This implies the existence of a non-vanishing area-form on $\cC$ which satisfies  $\iota_\ell \ve_{\cC}=0$,  where $\iota_{\ell}$ is the interior product. The value of $\ve_{\cC}$ at a point $x$ of $\cN$ is independent of the cut chosen that goes through $x$.
An adapted volume form $\ve_{\cN}$ on $\cN$ is defined to be such that one has $\iota_\ell \ve_{\cN}=\ve_{\cC}$. The expansion scalar $\theta$ associated to $\ell$ relates these two forms through $\rd \ve_{\cC}=\theta \ve_{\cN}$.

The Carrollian structure defines a canonical line bundle  on a base manifold $S$, such that the vertical lines in the kernel of the bundle projection map are along $\ell$, which defines the vertical direction. 
To decompose the tangent space of $\cN$ into a direct sum of vertical plus horizontal directions we need
to introduce an Ehresmann connection \cite{Ciambelli:2019lap}, that is, a 1-form $k=k_a \rd x^a$ dual to $\ell$, such that $\iota_\ell k=1$. This defines a notion of horizontality: a vector field $Y$ is horizontal on $\mathcal{N}$ if $\iota_Y k=0$.

Given that the metric is degenerate, we cannot invert it. Instead, we can define a symmetric horizontal tensor $q^{ab}$ such that its composition with the metric gives the horizontal projector $q_{ac}q^{cb}= \delta_a^b - k_a\ell^b= q_a{}^b$. This projector  satisfies $\ell^a q_a{}^b=0=q_a{}^b k_b$ and $q_a{}^bq_{bc}=q_{ac}$. 
Then, one has $Y^a q_a{}^b=Y^b$ for a horizontal vector. The Ehresmann connection $k$ allows us to decompose the volume form on $\cN$ as $\ve_{\mathcal{N}}=k\wedge \ve_{\cC}$. Moreover, we can harmlessly raise indices of horizontal tensors using $q^{ab}$. Indeed, given a generic tensor ${\cal A}_{ab}$ on $\cN$, we can uniquely define ${\cal A}_a{}^b= {\cal A}_{ac} q^{cb}$, which satisfies  ${\cal A}_a{}^c q_{cb}={\cal A}_{ab}$ and ${\cal A}_a{}^b k_b=0$. 

The set of data $(q_{ab}, \ell^b, k_a)$ defines a ruled Carrollian structure on $\mathcal{N}$. This structure is not unique: on top of the rescaling $(\ell,k) \rightarrow (\mathrm{e}^\lambda\ell, e^{-\lambda } k)$ mentioned above, one can shift the Ehresmann connection $k\rightarrow k-\zeta$ with $\zeta$ an horizontal form, i.e. $\zeta_a\ell^a=0$. The rescaling and shift transformations constitute the symmetries of a ruled Carrollian structure. These are studied in detail in Section \ref{symm}.

Given a choice of Ehresmann connection, one computes the  Carrollian acceleration $\varphi$ and vorticity $\varpi$ defined in \cite{Ciambelli:2019lap}, which are   horizontal 1-form and 2-form, i.e.  $\iota_\ell\varphi=0=\iota_\ell \varpi$,  given by
\beq\label{dk}
\rd k=\varphi\wedge k-\varpi.
\eeq
Frobenius theorem guarantees that the $1$-form $k$ defines a foliation of cuts if and only if $\rd k\wedge k=0$, which requires $\varpi=0$. In this case we can adapt coordinates to have that $k=e^\beta \rd V$, where the foliation leafs are at constant $V$. The acceleration is a horizontal form which represents the Lie transport of $k$, 
\be 
\cL_\ell k_a =-\varphi_a.
\ee
In the following, prior to any gauge fixing and unless explicitly stated, we consider Ehresmann connections with non-vanishing vorticity and acceleration.

Once endowed with an Ehresmann connection, we can concisely express the action of the Carrollian connection $D_a$ on the ruled Carrollian structure $(q_{ab},\ell^a, k_a)$. 
First, from eq.~\eqref{Dq} one has 
\be \label{Dq2}
D_a q_{bc}= - \theta_{ab}k_c-\theta_{ac}k_b.
\ee
Then, the action on the volume form defines a one form $\omega=\omega_a\rd x^a$ via
\be \label{boostf}
D_a \ve_{\mathcal{N}} = -\omega_a \ve_{\mathcal{N}}.
\ee
This one form also enters the covariant derivative of the Carrollian vector field $\ell$ as
\be \label{dl}
(D_a-\omega_a)\ell^b =\theta_a{}^b,
\ee 
where $\theta_a{}^b= \theta_{ac} q^{cb}$. This implies that $\omega_a = k_b D_a \ell^b$.
The $1$-form $\omega$ can be decomposed into a transverse and horizontal part as 
\beq\label{om}
\omega_a =\kappa k_a + \pi_a\,,
\eeq
where, when $\cN$ is embedded, $\pi_a$ is the H\'ajiček connection and $\kappa= k_b D_{\ell}\ell^b$ is the inaffinity of $\ell$ on $\cN$. 

Lastly, as discussed in \cite{Freidel:2022bai, Freidel:2024emv, Ciambelli:2025unn}, the Carrollian connection acts on $k_a$ in a highly non-trivial way, in order to be compatible with eqs.~(\ref{dk},\ref{dl}), and the torsionless condition. One has
\be\label{Dk}
(D_a+\omega_a)k_b =\bar\theta_{ab}-k_a(\pi_b+\varphi_b),
\ee
where $\bar\theta_{ab}$ is a horizontal tensor whose symmetric components are free  while its anti-symmetric part is given by the Carrollian vorticity, $\bar\theta_{[ab]}=-\frac12 \varpi_{ab}$. This shows in particular that the parallel transport of the Ehresmann connection along the null generator is given by 
\be \label{Dk1}
D_\ell k_a =-(\omega_a+\varphi_a).
\ee 
Therefore, the Carrollian connection depends on  the intrinsic Carrollian data $(\theta_{ab}, \varphi_a, \omega_{ab})$, and is determined by the choice of a one form and horizontal symmetric tensor $(\omega_a, \bar\theta_{(ab)})$. This is thoroughly discussed in the review \cite{Ciambelli:2025unn}, where it is also derived that the standard Carrollian affine symbols are thus given by
\be\label{affSym}
\Gamma^a_{bc}=\left((\pa_b+\omega_b) k_c+k_b(\pi_c+\varphi_c)-\overline\theta_{bc}\right)\ell^a+\frac{q^{ad}}{2}(\pa_bq_{cd}+\pa_cq_{bd}-\pa_dq_{bc}).
\ee

If we view the null hypersurface $\cN$ from a spacetime perspective, we can understand the Carrollian connection as being induced by projecting the  bulk Levi-Civita connection $\nabla_\mu $  along the  rigging projector $\Pi_\mu{}^\nu=\delta_\mu^\nu- \ell_\mu k^\nu$, where $\ell_\mu = g_{\mu\nu}\ell^\nu$ is the null normal form of $\cN$ and $k^\mu= g^{\mu\nu} k_\nu$ is the null rigging vector. Given a projected tensor $\cA_\mu{}^\nu=\Pi_\mu{}^{\alpha} T_{\alpha}{}^{\beta} \Pi_{\beta}{}^\nu$, one defines \cite{Mars:1993mj} (see also \cite{Chandrasekaran:2021hxc, Freidel:2022bai})
\be
D_a \cA_b{}^c = \Pi_a{}^{\alpha}\Pi_b{}^{\beta } \nabla_{\alpha}\cA_{\beta}{}^{\gamma} \Pi_{\gamma}{}^c.
\ee
This definition implies that $D_a$ preserves the rigging projector $D_a\Pi_b{}^c=0$. 

To recapitulate, we have here introduced the salient features of null geometries, in the language of a ruled Carrollian structure. The Carrollian connection has been introduced either from projecting the ambient space Levi-Civita connection, or by intrinsically requiring absence of torsion and metricity in the horizontal directions. The free connection coefficients are then $\omega_a$ and  $\bar\theta_{(ab)}$.

\subsection{Dynamics}\label{S22}

The expansion tensor can be decomposed in terms of the expansion $\theta$ and shear $\sigma_a{}^{b}$
\be \label{thet}
\theta_{a}{}^b= \frac{1}{2} \theta q_{a}{}^b + \sigma_{a}{}^b, \quad \text{with} \quad \sigma_a{}^a=0.
\ee 
The shear encodes the transverse spin-two gravitational data crossing the hypersurface and therefore contains information about radiation. Its quadratic contribution to the Raychaudhuri equation acts as a source of focusing. It is well-known that the projection of Einstein's equations along a null hypersurface leads to the Raychaudhuri \cite{PhysRev.98.1123, Sachs:1961zz} and Damour  \cite{damour1978black} equations, see also \cite{price1986membrane}. Here, we wish to derive them as the conservation laws of a stress tensor intrinsically defined on $\cN$.

The intrinsic null Brown-York stress tensor is
\be \label{Tby}
8\pi G\,T_a{}^b \coloneqq D_a\ell^b - \delta_a{}^b D_c\ell^c.
\ee 
From the ambient space construction, this was defined in \cite{Chandrasekaran:2021hxc}  using the Weingarten map $W_a{}^b=D_a\ell^b =\omega_a \ell^b +\theta_a{}^b$, with trace $D_a\ell^a=\theta + \kappa$, see \cite{Gourgoulhon:2005ng} for more details on the Weingarten map. One important consequence of this definition is that $\ell$ is an eigenvector of $T_a{}^b$, i.e., $\ell^a T_a{}^b=-\frac{\theta}{8\pi G} \ell^b$. Then, this stress tensor can be decomposed as
\beq\label{Thydro}
T_a{}^b =\tau_a \ell^b +\tau_a{}^b=\frac1{8\pi G} \left(-\theta k_a\ell^b-\mu q_a{}^b+\pi_a \ell^b +\sigma_a{}^b\right),
\eeq
where
\beqn\label{tau}
8\pi G\,\tau_a=  \pi_a-\theta k_a,\qquad
8\pi G\, \tau_a{}^{b}=\sigma_a{}^{b}-\mu q_a{}^{b}.
\eeqn
Here, we have introduced the surface tension $\mu=  \kappa +\frac{\theta}{2}$. In \cite{Ciambelli:2023mir, Ciambelli:2024swv}, we have seen that $\mu$ is canonically conjugate to $\Omega$, and is a crucial quantity in the phase space and its quantization. 

In \cite{Chandrasekaran:2021hxc,Freidel:2022bai,Freidel:2022vjq}, it was shown that the Einstein's equations on $\mathcal{N}$ are given by
\be \label{DT}
D_b T_a{}^b= T^{\mathsf{mat}}_{a\ell},
\ee  
where $T_{ab}^{\mathrm{mat}}$ is the bulk matter stress tensor. This generalizes to null hypersurfaces the  Brown-York result \cite{brown1993quasilocal}. The Raychaudhuri and Damour equations are derived projecting \eqref{DT} into its vertical and horizontal components
\bea
8\pi G\,\ell^bD_aT_b{}^a
&=&-(\cL_{\ell}+\theta)[\theta]+\mu\theta-\sigma_a{}^b\sigma_b{}^a-R_{\ell\ell} \\
8\pi G\,q_a{}^bD_cT_b{}^c
&=& \left({\cal L}_\ell+\theta\right)[\pi_a] -\overline{D}_a\mu +(\theta-\mu)\varphi_a + (\overline{D}_b+\varphi_b)\sigma_{a}{}^b
-q_a{}^bR_{b\ell} \label{Damour0}
\eea
where $R_{ab}$ is the Ricci tensor, $R_{a\ell}=R_{ab} \ell^b$, and $\overline{D}_b$ denotes the horizontal derivative. For a generic tensor $X_{a_1\dots a_n}{}^{b_1\dots b_m}$,
\beq
\overline{D}_e X_{a_1\dots a_n}{}^{b_1\dots b_m}= q_e{}^f  q_{a_1}{}^{c_1}\cdots q_{a_n}{}^{c_n}q_{d_1}{}^{b_1}\cdots q_{d_n}{}^{b_n} D_f X_{c_1\dots c_n}{}^{d_1\dots d_m}\,.
\eeq
The derivation of Damour equation is given in appendix \ref{AppA}. Note, the vorticity $\varpi$ does not contribute to either equations, while the acceleration $\varphi$ contributes only to the Damour equation. 

In \cite{Ciambelli:2023mir, Ciambelli:2024swv}, and in the rest of this manuscript, we are interested in the densitized constraints
\bea
C= 8\pi G\,\Omega\, \ell^bD_aT_b{}^a,
\label{RayDamour}\qquad J_a=8\pi G\,\Omega q_a{}^bD_cT_b{}^c\,,
\eea
where $\Omega$ is the scalar area element such that $\ve_{\cC}= \Omega \rd^2 \sigma$, see \eqref{uni}. This makes  all the field-dependency explicit.

\subsection{Symmetries}\label{symm}

There are three symmetries involved in the definition of a ruled Carrollian structure: the rescaling of $\ell$, the shift of $k$, and the diffeomorphisms of the null manifold $\xi\in T\cN$. In preparation of the covariant phase space analysis in the next section, we describe the action of these symmetries using field space calculus. For a symmetry generator $\alpha$, we denote  $\hat\alpha$ its associated vector field action on fields and $\fL_{\hat{\alpha}}=I_{\hat{\alpha}}\delta+\delta I_{\hat\alpha}$ the field space Lie derivative, with $\delta$ the field variation, and $I_{\hat\alpha}$ the field space interior product. Details of this  notation introduced in \cite{Donnelly:2016auv} can be found e.g. in  \cite{Ciambelli:2022vot}.

\paragraph{Diffeomorphisms.}

The geometric structure is covariant under diffeomorphisms of the null hypersurface. The fields transform via the spacetime Lie derivative\footnote{We can replace the partial derivative with the Carrollian covariant derivative because the latter is torsionless.}
\beq
\fL_{\hat{\xi}} q_{ab} &=&\xi^c D_c q_{ab}+q_{ac}D_b \xi^c+q_{cb}D_a\xi^c\\
\fL_{\hat{\xi}}\ell^a &=& \xi^b D_b\ell^a-\ell^b D_b \xi^a\\
\fL_{\hat{\xi}}k_a &=& \xi^b D_b k_a+k_b D_a \xi^b \label{xik}\\
\fL_{\hat{\xi}} \tau^{ab} &=&\xi^c D_c \tau^{ab}-\tau^{ac}D_b \xi^c-\tau^{cb}D_a\xi^c\\
\fL_{\hat{\xi}}\tau_a &=& \xi^b D_b\tau_a+\tau_b D_a \xi^b.
\eeq
Both the stress tensor and its divergence transform covariantly under diffeomorphisms.

\paragraph{Rescaling.} While we referred to this symmetry in \cite{Ciambelli:2023mir} as ``internal local Carroll boost'', we call it ``rescaling'' in this manuscript to avoid confusion with previous literature, where the term Carroll boost is used to talk about shifts in $k$, see, e.g., \cite{deBoer:2017ing, Armas:2023dcz} and references therein. Rescalings of $\ell$ lead to the same Carrollian structure, since the condition $\ell^aq_{ab}=0$ is unaffected. Then, to preserve $\ell^ak_a=1$, we must transform $k$ as well. From this we can derive the transformation rules
\beq\label{lambdak}
\fL_{\hat\lambda} \ell^a=\lambda \ell^a,\quad \fL_{\hat\lambda} q_{ab}=0, \quad \fL_{\hat\lambda} k_a=-\lambda k_a,\quad  \fL_{\hat\lambda} \ve_{\cN}=-\lambda \ve_{\cN} \quad \fL_{\hat\lambda}\theta=\lambda \theta.
\eeq

From an intrinsic standpoint, the action of this symmetry on the independent data parametrizing the affine symbols \eqref{affSym} must be prescribed.\footnote{From the extrinsic viewpoint, this symmetry corresponds to the rescaling of $f$ in \cite{Odak:2023pga}.} Given that this symmetry leaves completely unaffected the vertical/horizontal split, and it does not act on the metric, it is natural to enforce that
\be\label{lambdaSym}
[\fL_{\hat\lambda},D_a]=0,
\ee 
that is, $\fL_{\hat\lambda}\Gamma^a_{bc}=0$.
Given $\varphi_a=-\cL_{\ell}k_a$, one can derive $\fL_{\hat\lambda}\varphi_a=-\overline{D}_a\lambda$. Acting on \eqref{dl} with $k_b$, we get $k_b(D_a-\omega_a)\ell^b =0$. From this, thanks to \eqref{lambda}, we get
\beq
\fL_{\hat\lambda}\omega_a=D_a\lambda.
\eeq
Given then that $\kappa= \ell^a \omega_a $, we then have
\beq
\fL_{\hat\lambda}\kappa=\lambda\kappa+D_{\ell}\lambda.
\eeq
The action of this symmetry on $\pi_a=q_a{}^b\omega_b$ is therefore
\beq
\fL_{\hat\lambda}\pi_a=\oD_a\lambda
\eeq
and we see that the combination $\pi_a+\varphi_a$ is invariant under rescaling. From this last fact, $\fL_{\hat\lambda}\Gamma^a_{bc}=0$ implies $\fL_{\hat\lambda}\overline\theta_{ab}=-\lambda\overline\theta_{ab}$.

To have all the ingredients to derive the transformation of the stress tensor, we need to study how the shear transforms. We first obtain from \eqref{Dq}
\beq
\fL_{\hat\lambda}\theta_{ab}=\lambda \theta_{ab}.
\eeq
Then, given that $\fL_{\hat\lambda}q_{a}{}^{b}=0=\fL_{\hat\lambda}q^{ab}$, this implies $\fL_{\hat\lambda}\theta_{a}{}^{b}=\lambda\theta_{a}{}^{b}$, and thus
\beq
\fL_{\hat\lambda}\sigma_{a}{}^{b}=\lambda\sigma_{a}{}^{b}.
\eeq

Eventually, we can compute
\beq\label{lamtau}
\fL_{\hat\lambda}\tau_{a}{}^{b}=\lambda\tau_{a}{}^{b}-\frac{D_\ell\lambda}{8\pi G}q_{a}{}^{b}\qquad \fL_{\hat\lambda}\tau_a=\frac{\oD_a \lambda}{8\pi G},
\eeq
which implies $\fL_{\hat\lambda}\tau^{ab}=\lambda\tau^{ab}-\frac{D_\ell\lambda}{8\pi G}q^{ab}$, a useful equation for the phase space analysis. We thus have
\beq
\fL_{\hat\lambda}T_b{}^a=\lambda T_b{}^a+\frac1{8\pi G}\Big(\ell^a D_b\lambda-\delta_b{}^a D_\ell\lambda\Big).
\eeq
An important crosscheck of this follows straightforwardly from the action of $\fL_{\hat\lambda}$ on eq.~\eqref{Tby}, and it implies $\fL_{\hat\lambda}(\ell^b T_b{}^a)=-\frac{\lambda\theta}{4\pi G}\ell^a$.

From these equations, using again \eqref{lambdaSym} and $D_{[a}D_{b]}\lambda=0$, we finally obtain
\beq
\fL_{\hat\lambda}D_aT_b{}^a=\lambda D_aT_b{}^a,
\eeq
demonstrating that the equations of motion \eqref{DT} are covariant under the rescaling symmetry. We finish the list of rescaling transformations displaying its action on the Raychaudhuri and Damour constraints as defined in eq.~\eqref{RayDamour}
\beq\label{lamCJ}
 \fL_{\hat\lambda}C = 2\lambda C\qquad  \fL_{\hat\lambda}J_a=\lambda J_a.
\eeq
Therefore, the Raychaudhuri constraint has rescaling weight $2$ while the Damour constraint has weight $1$, and as such, we expect that it will play the role of a CFT current in the quantization.
Note that the covariance of the Damour equation under rescaling can be  proven directly using the Carrollian commutation relations 
\be 
[\ell,\overline\pa_a] = -\pa_a\ell^b \overline\pa_b +\varphi_a \ell, \qquad [\overline\pa_a,\overline\pa_b]= \varpi_{ab}\ell.
\ee

\paragraph{Shift.} 

As we mentioned, there is another symmetry involved in specifying a ruled Carrollian structure: any shift in $k$ that is orthogonal to $\ell$ leaves its defining equation $\iota_\ell k=1$ unaltered. This is realized on the field space introducing an horizontal $1$-form $\zeta$ satisfying $\iota_\ell\zeta=0$. It acts on the fields as
\beq\label{zetak}
\fL_{\hat\zeta} \ell^a=0, \quad \fL_{\hat\zeta} k_a=-\zeta_a,\quad  \fL_{\hat\zeta} \ve_{\cN}=0 \quad \fL_{\hat\zeta} q_{ab}=0,\quad \fL_{\hat\zeta}\theta=0,\quad \fL_{\hat\zeta} q_{a}{}^b=\zeta_a \ell^b,
\eeq
where we used that $\fL_{\hat\zeta} \ve_{\cN}=\fL_{\hat\zeta} (k\wedge \ve_{\cC})=-\zeta\wedge \ve_{\cC}=0$ since both $\zeta$ and $\ve_{\cC}$ are orthogonal to $\ell$. Then, using that $\varphi_a=-\cL_{\ell}k_a$, we obtain 
\beq
\fL_{\hat\zeta}\varphi_a=\cL_{\ell}\zeta_a,
\eeq
and we remark that this is horizontal, that is, $\ell^a \fL_{\hat\zeta}\varphi_a=0$. For the vorticity, using eq.~\eqref{dk}, we obtain
$\fL_{\hat\zeta} \varpi_{ab}= (\overline{D}_a-\varphi_a)\zeta_b - (\overline{D}_b-\varphi_b)\zeta_a$. Furthermore, since $\fL_{\hat\zeta}\theta_{ab}=\frac12 \fL_{\hat\zeta}(\cL_{\ell}q_{ab})=0$, one has 
\beq\label{zetatheta}
\fL_{\hat\zeta}\theta_a{}^b=\theta_a{}^c\zeta_c\ell^b,\qquad \fL_{\hat\zeta}\theta^{ab}=\zeta_c\left(\theta^{ac}\ell^b+\theta^{cb}\ell^a\right).
\eeq

To proceed, from an intrinsic standpoint, we need to postulate how this symmetry acts on the connection. While for the rescalings we postulated that the connection was invariant, we here require a non-trivial action on the connection symbols, motivated by the fact that this symmetry shifts the horizontal subbundle.  Such a transformation rule is dictated by the bulk interpretation of the shift symmetry as changes in the rigging structure \cite{Odak:2023pga}, and the fact that the connection is not required to be invariant, but instead its transformation can be non-linear, to enforce this is a symmetry of the equations of motion. Under a shift of the rigging, while keeping the ambient Levi–Civita connection fixed, the induced Carrollian connection transforms as
\beq\label{zetacri}
\fL_{\hat\zeta}\Gamma^c_{ab}=-\theta_{ab}\zeta^c\,,
\eeq
where $\zeta^a$ satisfies $\zeta^a q_{ab}=\zeta_b$ and $\zeta^a k_a=0$. From the intrinsic viewpoint, this is a postulate. Such a transformation preserves the  condition \eqref{Dq2} since 
\bea
\fL_{\hat\zeta} (D_a q_{bc}) = - 2\fL_{\hat\zeta}\Gamma_{a(b}^d q_{c) d} =
2 \theta_{a(b}\zeta_{c)} = - \fL_{\hat\zeta}( \theta_{ab}k_c+ \theta_{ac}k_b).
\eea

From these relations, we can eventually derive how the shift symmetry acts on $\kappa$ and $\pi_a$
\beq
\fL_{\hat\zeta}\omega_a=-\zeta_b {D}_a \ell^b ,\quad \fL_{\hat\zeta}\kappa=0,\quad \fL_{\hat\zeta}\pi_a=\kappa\zeta_a-{D}_a \ell^b \zeta_b=(D_\ell+\kappa)\zeta_a-\cL_\ell \zeta_a.
\eeq
Therefore, the combination $\pi_a+\varphi_a$, which is invariant under rescaling, simply  transforms as 
\be \label{piphi}
\fL_{\hat\zeta}(\pi_a+\varphi_a) =(D_\ell+\kappa)\zeta_a.
\ee 

Moreover, from eq.~\eqref{zetatheta}, using that $\fL_{\hat\zeta}q^{ab}=\zeta^a\ell^b+\zeta^b\ell^a$, we gather
\beq
\fL_{\hat\zeta}\sigma^{ab}=\zeta_c \left(\sigma^{ac}\ell^b+\theta^{cb}\ell^a\right).
\eeq
Using this and eq.~\eqref{piphi}, one finds
\beq\label{zetatau}
\fL_{\hat\zeta}\tau^{ab}=\zeta_c\left(\tau^{cb}\ell^a+\tau^{ca}\ell^b\right),\quad\fL_{\hat\zeta}\tau_a=\frac1{8\pi G}\left((\theta+\kappa)\zeta_a-D_a \ell^b \zeta_b \right)=-\tau_a{}^b\zeta_b.
\eeq
From this, we obtain that the Carrollian stress tensor is invariant under the shift symmetry
\beq
\fL_{\hat\zeta}T_a{}^b=0.
\eeq
This can be checked directly as 
\be
\fL_{\hat\zeta}T_a{}^b= \fL_{\hat\zeta}\Gamma_{ac}^b \ell^c - \delta_a{}^b \fL_{\hat\zeta}\Gamma_{ac}^c \ell^a =(\zeta^b \theta_{ab} - \zeta^c \theta_{ca})\ell^a=0.
\ee
This is the first important result of this manuscript. We prescribed the correct set of properties of the shift symmetry to ensure that the null Brown-York stress tensor is completely invariant. Ultimately, this was possible thanks to the requirement \eqref{zetacri}, which a posteriori motivates this choice. The second important result is that the equations of motion are also shift invariant. Indeed
\beqn
\fL_{\hat\zeta}D_aT_b{}^a&=&\fL_{\hat\zeta}\Gamma_{ac}^aT_b{}^c-\fL_{\hat\zeta}\Gamma_{ab}^cT_c{}^a\\
&=&-\theta_{ac}\zeta^aT_b{}^c+\theta_{ab}\zeta^cT_c{}^a\\
&=&-\frac{\zeta^a}{8\pi G}\left(\theta_{ac}\sigma_b{}^c-\theta_{cb}\sigma_a{}^c\right)\\
&=&-\frac{\zeta^a}{8\pi G}\left(\sigma_{ac}\sigma_b{}^c-\sigma_{cb}\sigma_a{}^c\right)\\
&=&0.
\eeqn

From this, we read the action of the shift symmetry on the constraints \eqref{RayDamour},
\bea\label{zetCJ}
\fL_{\hat\zeta} C &=& 8\pi G\,\fL_{\hat\zeta}\left(\Omega\, \ell^bD_aT_b{}^a\right)=0,\cr
\fL_{\hat\zeta}J_a &=& 8\pi G\,\Omega \fL_{\hat\zeta}q_a{}^b\,D_cT_b{}^c=8\pi G\,\Omega \zeta_a\ell^b\,D_cT_b{}^c=\zeta_a C\,.
\eea
The rescaling \eqref{lamCJ} and shift \eqref{zetCJ} action on the constraints can be combined in the suggestive form
\beqn
\fL_{(\hat\lambda,\hat\zeta)}\begin{pmatrix}C\cr J_a\end{pmatrix}
=\begin{pmatrix}2\lambda & 0\cr \zeta_a & \lambda q_a{}^b\end{pmatrix}\begin{pmatrix}C\cr J_b\end{pmatrix}.
\eeqn
Therefore, the action is lower triangular. Since the Raychaudhuri and Damour constraints generate time reparameterization and spatial transformations, respectively, this is consistent with the semi-direct product structure $\text{Diff}(\cC)\ltimes \RR^\cC$.
 
So in conclusion, we have discussed $3$ different symmetries, for which the equations of motion are covariant. 
While the connection is invariant under rescaling, for the shifts we required it to transform inhomogeneously, as it happens for diffeomorphisms. This transformation ensures that both the stress tensor and the intrinsic null equations are invariant under shifts.

\section{Symplectic Analysis}\label{s3}

We now turn our attention to the symplectic structure of gravity on a null hypersurface. After reviewing the action of the rescaling symmetry and diffeomorphisms, we demonstrate that the shift symmetry is pure gauge. We then use this fact to promote the Ehresmann connection $k$ to a background structure, without restricting the diffeomorphisms on $\cN$. This is achieved combining diffeomorphisms with rescaling and shift. 

\subsection{Symplectic Potential and Charges}

With judiciously chosen corner terms, the gravitational presymplectic potential induced on a null hypersurface is given by \cite{Chandrasekaran:2021hxc, Ciambelli:2023mir,Chandrasekaran:2023vzb}\footnote{As discussed in \cite{Odak:2023pga}, the choice of corner term and polarization used to reach $\Theta^{\mathsf{can}}$ leads to covariant phase space anomalies. While it would be interesting to perform the prime phase space analysis using the formalism of \cite{Odak:2023pga}, we proceed with this potential, since ultimately we are interested in the bulk Poisson brackets, which are independent from $\delta-$ and $\rd-$exact terms in $\Theta$.}
\beq\label{thcan}
\Theta^{\mathsf{can}}=\int_{\cN}\theta^{\mathsf{can}}=\int_{\cN} \ve_{\cN}\Big(\frac{1}{2}
\tau^{ab}\delta q_{ab}
-\tau_a\delta \ell^a\Big).
\eeq
For a generic field space morphism $\hat\alpha$, using the field-space notations introduced in the previous section, we can compute the Noether current via the formula
\beq
I_{\hat\alpha}\Omega^{\mathsf{can}}=I_{\hat\alpha}(\delta\Theta^{\mathsf{can}})=\fL_{\hat\alpha}\Theta^{\mathsf{can}}-\delta I_{\hat\alpha}\Theta^{\mathsf{can}}.
\eeq

Choosing $\hat\alpha$ to be a diffeomorphism, we find $\fL_{\hat{\xi}}\Theta^{\mathsf{can}}= \cF^{\mathsf{can}}_\xi$, where the canonical flux $\cF^{\mathsf{can}}_\xi$  is a corner term given by 
\be \label{fff}
\cF^{\mathsf{can}}_\xi= \int_{\cC} \Big(\frac{1}{2}\tau^{ab}\delta q_{ab}-\tau_a\delta \ell^a\Big) \iota_{\xi}\ve_{\cN}.
\ee
The charge is then extracted from
\be \label{noexi}
I_{\hat{\xi}}\Omega^{\mathsf{can}} = -\delta Q_\xi+\cF_\xi^{\mathsf{can}},
\ee
and explicitly given by
\bea\label{ixcan}
Q_\xi=I_{\hat\xi}\Theta^{\mathsf{can}}=
\int_{\cN} \Big(T_a{}^b D_b\xi^{a}-\tau_a \xi^b D_b\ell^a \Big) \ve_{\cN} =
-\int_{\cN} \xi^{a} D_bT_a{}^b \ve_{\cN} 
+  \int_{\cC} \xi^{a}T_a{}^b \ve_{b}.  
\eea
In this expression, we introduced $\ve_a=n_a \ve_{\cC}$, where $n_a$ is the normal $1$-form to the cut $\cC$. At this stage, we are keeping the cut generic. On shell of the constraints (signaled by $\hat{=}$), we obtain
\be \label{charg}
Q_\xi \,\hat{=}\,
 \int_{\cC} \xi^aT_a{}^b \ve_b.  
\ee 
The result \eqref{ixcan} can be regarded as the reason why the stress tensor \eqref{Thydro} is the null analogue of the Brown-York stress tensor.

We then focus on the rescaling. In the next section, we introduce the concept of prime phase space, which is reached by constraining the rescaling $\lambda$ as a function of the phase space variables. In preparation of that analysis, we therefore compute the charges assuming that $\hat\lambda$ is field dependent, such that $\delta\lambda\neq 0$. Using eqs.~\eqref{lambdak} and \eqref{lamtau}, we thus get
\beq
\fL_{\hat\lambda}\Theta^{\mathsf{can}}&=&-\frac1{8\pi G}\int_{\cN}\ve_{\cN}\left(\frac12 \ell(\lambda)q^{ab}\delta q_{ab}+\olD_a\lambda\delta \ell^a-\theta\delta\lambda\right)\\&=&-\frac1{8\pi G}\int_{\cN}\delta\left(\ve_{\cN}\ell(\lambda)\right)+\frac{1}{8\pi G}\int_{\cN}\ve_{\cN}(\ell+\theta)(\delta\lambda),
\eeq
where we used $\delta \ve_{\cN} = (\frac12 q^{ab}\delta q_{ab} - k_a \delta \ell^a)\ve_{\cN}$. Computing then
\beq
I_{\hat\lambda}\Theta^{\mathsf{can}}=\frac1{8\pi G}\int_{\cN}\ve_{\cN}\lambda \theta,
\eeq
we obtain
\beq\label{ilam}
I_{\hat\lambda}\Omega^{\mathsf{can}}=\fL_{\hat\lambda}\Theta^{\mathsf{can}}  -\delta I_{\hat\lambda}\Theta^{\mathsf{can}}=-\frac1{8\pi G}\int_{\cN}\left(\delta\left(\ve_{\cN}(\ell+\theta)(\lambda)\right)-\ve_{\cN}(\ell+\theta)(\delta\lambda)\right).
\eeq

To reduce this to an integral over a cut $\cC$, we note that for a generic vector $\cA^a$, one has
\beq\label{stok}
\int_{\cC}\ve_a\cA^a=\int_{\cN}D_a(\ve_{\cN}\cA^a)=\int_{\cN}\ve_{\cN}(D_a-\omega_a)\cA^a.
\eeq
where we used \eqref{boostf}.
Suppose then that $\cA^a=\chi \ell^a$ for a scalar $\chi$. Using eqs.~\eqref{boostf} and \eqref{om}, one gets
\beq\label{stoke}
\int_{\cC}\ve_a \ell^a \chi =\int_{\cN}\ve_{\cN}\left(\ell(\chi)+\chi (D_a\ell^a-\kappa)  \right)=\int_{\cN}\ve_{\cN}\left(\ell+\theta\right)(\chi),
\eeq
where we used that $D_a\ell^a=\kappa+\theta$ (see eq.~\eqref{dl}) and $\ell^a \ve_a = \ve_\cC$.

Using this fact, we can process eq.~\eqref{ilam} into
\beq\label{ilam2}
I_{\hat\lambda}\Omega^{\mathsf{can}}=- \frac1{8\pi G}\int_{\cC}\delta\left( \ell^a\ve_{a}\,\lambda \right)+\frac1{8\pi G}\int_{\cC}\ell^a\ve_{a}\delta\lambda.
\eeq
When $\delta\lambda=0$, this is the result derived in \cite{Ciambelli:2023mir}, showing that the rescaling charge is given by the corner area element
\beq\label{Gl}
Q_\lambda =\frac1{8\pi G}\int_{\cC}\ell^a\ve_{a}\, \lambda \,,
\eeq
where we used that in the absence of canonical fluxes,   $I_{\hat\lambda}\Omega^{\mathsf{can}}=-\delta Q_\lambda$.
We observe that the cut $\ve_{\cC}$ is not necessarily a constant-$v$ cut. In case it is not, its form in local coordinates depends on the derivative of the function defining the cut's position. 
Reintroducing the field-dependency of $\lambda$, we can eventually recast eq.~\eqref{ilam2} as
\beq\label{Noe}
I_{\hat\lambda}\Omega^{\mathsf{can}}=-\delta Q_{\lambda}+Q_{\delta\lambda}.
\eeq

Finally, we focus our attention on the shift. Choosing $\hat\alpha$ to be a shift, eqs.~\eqref{zetak} and \eqref{zetatau} give
\beq
\fL_{\hat\zeta}\Theta^{\mathsf{can}}=\int_{\cN}\ve_{\cN}\left(\frac12\zeta_c(\tau^{cb}\ell^a+\tau^{ca}\ell^b)\delta q_{ab}+\tau_a{}^b \zeta_b \delta\ell^a\right)=\int_{\cN}\ve_{\cN}\left(-\zeta_c\tau^{cb}\delta \ell^a q_{ab}+\tau_a{}^b \zeta_b \delta\ell^a\right) =0.
\eeq
This, together with $I_{\hat\zeta}\Theta^{\mathsf{can}}=0$, proves the main result of this section
\beq
I_{\hat\zeta}\Omega^{\mathsf{can}}=0.
\eeq
Therefore, the shift symmetry is pure gauge, as already anticipated in \cite{Freidel:2024emv}. Stated differently, the Ehresmann connection carries no independent bulk phase-space degree of freedom and may be fixed as part of the background structure.
To recap, we have shown that diffeomorphisms lead to non-vanishing Noether charges and non-vanishing symplectic fluxes, rescalings lead to non-vanishing integrable Noether charges -- plus the standard $Q_{\delta\lambda}$ term -- that localize to corners off-shell and do not possess fluxes, whereas shifts are completely pure gauge symmetries with vanishing Noether charges off-shell.

Since we are interested in describing the constraints and the associated phase space in the bulk of the null hypersurface, we can systematically exploit rescaling and shifts to judiciously fix a geometric background structure. This is the essence of the prime phase space analysis hereafter.

\subsection{Prime Phase Space}

We now improve the prime phase space analysis of \cite{Ciambelli:2023mir} to include the shift contribution. 

The reason for restricting to the prime phase space is to dispose of a term in the symplectic structure proportional to $k_a\delta\ell^a$, as this would introduce a new symplectic datum on the hypersurface coupled to the spin-0 sector. In \cite{Ciambelli:2023mir}, we imposed the stringent condition that $\delta\ell^a$ is proportional to $\ell^a$, as this allowed us to also decouple the spin-1 sector and focus only on the Raychaudhuri constraint. Here, we want to include the Damour constraint in our analysis, and thus the horizontal spin-1 sector, which is achieved imposing the less stringent condition
\beq\label{deltak}
\delta k_a=0.
\eeq
Indeed, this maintains the horizontal spin-1 symplectic data in our phase space while it discards the term $k_a\delta\ell^a$. This condition means that we treat $k$ as a background datum. Since $k$ is pure gauge it is possible to treat the Ehresmann connection as background and fix it to a particular choice. This is what  we implement in the rest of the paper.  

This condition can be imposed in a fully diffeomorphism-covariant manner, by exploiting the rescaling and shift symmetries. For that, consider the combined action of diffeomorphism \eqref{xik} with rescaling \eqref{lambdak} and shift  \eqref{zetak}. For the diffeomorphism action, we decompose $\xi=f\ell+Y$ with $Y^aq_a{}^b=Y^b$ and use \eqref{dk} to rewrite
\be
\fL_{\hat\xi}k=\rd \iota_\xi k+\iota_\xi \rd k=\rd \iota_\xi k+\iota_\xi (\varphi\wedge k-\varpi)=\rd f+Y^a\varphi_a k-f \varphi-\iota_Y \varpi,
\ee
where we used that both $\varphi$ and $\varpi$ are horizontal forms. Therefore, we gather
\be
\fL_{(\hat \xi,\hat \lambda,\hat\zeta)}k_a=(D_a -\varphi_a)f +Y^b \varphi_b k_a-Y^b \varpi_{ba}-\zeta_a-\lambda k_a.
\ee
Compatibility with \eqref{deltak} requires we set this expression to zero, and solve for $\lambda$ and $\zeta$. We do so projecting it into horizontal and vertical parts and choosing the gauge parameter to be dependent on $\xi$ through\footnote{In \cite{Ciambelli:2023mir}, we focused on time reparameterization, and implicitly imposed $\varphi=0$. The prime condition was $\lambda=\ell(f)$.}
\beq\label{lamxi}
\lambda(\xi) =\ell(f)+Y^b\varphi_b,&\label{lambda}\qquad 
\zeta_a(\xi)=(\overline{D}_a-\varphi_a) f -Y^b \varpi_{ba}.
\eeq
We thus reach the prime phase space conditions, fixing $\lambda=\lambda(\xi)$ and $\zeta=\zeta(\xi)$ such that $\fL_{(\hat \xi,\hat \lambda,\hat\zeta)}k=0$. 

Since from now on $\lambda$ and $\zeta$ are constrained in this manner, we refer to the prime Lie derivative as $\fL'_{\hat \xi}$. Its action on the geometric fields is
\beq
\fL'_{\hat \xi} q_{ab} &=&\xi^c D_c q_{ab}+q_{ac}D_b \xi^c+q_{cb}D_a\xi^c,\\
\fL'_{\hat \xi} \ell^a &=& \xi^b D_b\ell^a-\ell^b D_b \xi^a+(\ell(f)+Y^b\varphi_b)\ell^a,\label{Lpell}\\
\fL'_{\hat \xi} k_a &=& 0.
\eeq
Given that $k_a\delta \ell^a=0$, we must satisfy $0=k_a\fL'_{\hat \xi} \ell^a =k_a\cL_{\xi} \ell^a+\lambda(\xi)$. Indeed, we have that 
\beq
k_a\cL_{\xi}\ell^a=k_a(\xi^bD_b\ell^a-\ell^bD_b\xi^a)=-\ell^a\xi^bD_bk_a-D_\ell(k_a\xi^a)+\xi^aD_\ell k_a,
\eeq 
then, using eqs.~\eqref{Dk} and \eqref{Dk1}, we gather
\beq
k_a\cL_{\xi}\ell^a=-\ell(f)-Y^a\varphi_a=-\lambda(\xi),
\eeq
as desired.

This implies that $\fL'_{\hat \xi} \ell^a$ is horizontal, which can also be derived by direct inspection
\beq
\fL'_{\hat \xi} \ell^a & =& \xi^b D_b\ell^a-\ell^b D_b \xi^a+(\ell(f)+Y^b\varphi_b)\ell^a\\
&=&Y^b(D_b+ \varphi_b) \ell^a-D_\ell Y^a \\ 
&=&Y^b\theta_b{}^a+ \ell^a Y^b(\pi_b+\varphi_b)-D_\ell Y^a \\ 
&=& Y^b \theta_b{}^a -q_b{}^a D_\ell Y^b,
\eeq
where in the third equality we used eq.~\eqref{dl}, while in the fourth we used eq.~\eqref{Dk1} and the fact that $\ell^a Y^b D_\ell k_b=-\ell^a k_bD_\ell Y^b=q_b{}^a D_\ell Y^b-D_\ell Y^a$.

Since the shift charge is zero, we can compute the prime phase space charges summing together the contribution from the diffeomorphism and the rescaling. From eqs.~\eqref{noexi} and \eqref{Noe}, the total contraction of the symplectic 2-form is
\beq\label{noeprime}
I'_{\hat\xi}\Omega^{\mathsf{can}}=-\delta\left(Q_{\xi}
+Q_{\lambda(\xi)}\right)+\cF_\xi^{\mathsf{can}}+Q_{\delta\lambda(\xi)}.
\eeq

Using \eqref{ixcan} and \eqref{Gl}, and decomposing the symmetry generators as $\xi=f\ell+Y$ in \eqref{ixcan}, we confirm for the vertical part ($Y^a=0$) the result obtained in \cite{Ciambelli:2023mir}, namely,
\beqn
M_f'&=&Q_{f\ell}+Q_{\lambda(f)}=
\frac{1}{8\pi G}\int_{\cN} \ve_{\cN}\Big(
f \sigma_a{}^b\sigma_b{}^a -\theta(\ell+\mu)(f)+(\ell+\theta)\lambda(f)\Big)
\\
&=&
-\frac{1}{8\pi G}\int_{\cN} \ve^{(0)}_{\cN}f C 
+\frac{1}{8\pi G}\int_{\cC}\ell^a\ve_{a} (\ell-\theta)(f),  
\eeqn
where we introduced the notation $\ve_{\cN}=\Omega \ve^{(0)}_{\cN}$, see \eqref{uni}. This is needed, as the constraints \eqref{RayDamour} are densities. For the horizontal part ($f=0$), the computation is more involved. Using again \eqref{ixcan} and \eqref{Gl}, we first get
\beqn
P_Y'=Q_{Y}+Q_{\lambda(Y)}=\int_{\cN} \ve_{\cN}\Big(
T_a{}^b D_bY^a-\tau_a Y^bD_b\ell^a+\frac{1}{8\pi G}(\ell+\theta)\lambda(Y)
\Big)\,.
\eeqn
Then, given \eqref{tau} and \eqref{lamxi}, we obtain
\beqn\label{Pyp}
P_Y'=\int_{\cN} \ve_{\cN}\Big(
\tau_a{}^b D_bY^a+\tau_a\cL_{\ell}Y^a+\frac{1}{8\pi G}(\ell+\theta)(Y^b\varphi_b)
\Big)\,.
\eeqn

Now, in order to use \eqref{stok}, we rewrite the first term as
\beq\label{step1}
\tau_a{}^b D_b Y^a=
(D_b-\omega_b)\big(\tau_a{}^b Y^a\big)-Y^aD_b\tau_a{}^b+\pi_b\tau_a{}^b Y^a,
\eeq
where we used that $\omega_b\tau_a{}^b=\pi_b\tau_a{}^b$.
In order to isolate the Damour constraint, we further need to process the second term in \eqref{step1}. Using the results derived in Appendix \ref{AppA}, and in particular eq. \eqref{tauq}, we get
\be\label{T1}
-Y^aD_b\tau_a{}^b=-Y^a(\overline{D}_b+\varphi_b)\tau_a{}^b-Y^a\tau_a{}^b\pi_b\,.
\ee
This combines with \eqref{step1} to give
\beq\label{step2}
\tau_a{}^b D_b Y^a=
(D_b-\omega_b)\big(\tau_a{}^bY^a\big)-Y^a(\overline{D}_b+\varphi_b)\tau_a{}^b.
\eeq

Focus now on the second term in \eqref{Pyp}, using $\tau_aY^a=\frac{\pi_aY^a}{8\pi G}$ and $\varphi_a=-\cL_{\ell}k_a$, we can process it into
\beqn\label{T2}
\tau_a\cL_{\ell}Y^a&=&\frac{1}{8\pi G}\left(\ell(\pi_aY^a)-Y^a\cL_{\ell}(-\theta k_a+\pi_a)\right)\\
&=&\frac{1}{8\pi G}\left(\ell(\pi_aY^a)-Y^a\varphi_a\theta-Y^a\cL_{\ell}\pi_a\right)\\
&=&\frac{1}{8\pi G}\left((\ell+\theta)(\pi_aY^a)-Y^a\varphi_a\theta-Y^a(\cL_{\ell}+\theta)\pi_a\right)\,.
\eeqn

Eventually, combining \eqref{T1} and \eqref{T2} into \eqref{Pyp}, and using \eqref{tau}, we reach
\beqn
P_Y'&=&\frac{1}{8\pi G}\int_{\cN} \ve_{\cN}\Big(
-Y^a\big((\cL_\ell+\theta)\pi_a+\theta\varphi_a+(\overline{D}_b+\varphi_b)(\sigma_a{}^b-\mu q_a{}^b)\big)\\
&&+(D_b-\omega_b)\big((\sigma_a{}^b-\mu q_a{}^b)Y^a\big)+\frac{1}{8\pi G}(\ell+\theta)(Y^b(\pi_b+\varphi_b))\Big)\,.
\eeqn
The first line is exactly the Damour constraint, as written in \eqref{DamA}, while the second line is a total derivative, thanks to \eqref{stok} and \eqref{stoke}. Therefore, we obtain the final expression
\beqn
P_Y'=-\frac{1}{8\pi G}\int_{\cN} \ve^{(0)}_{\cN}
Y^aJ_a+\frac{1}{8\pi G}\int_{\cC} Y^b\left((\pi_b+\varphi_b)\ell^a+\sigma_b{}^a -\mu q_b{}^a\right)\ve_a\,.
\eeqn

We remark that, on shell, and for a cut adapted to $\ell$, we have
\be
M_f'\hat{=}\frac{1}{8\pi G}\int_{\cC}\ve_{\cC} (\ell-\theta)(f)=\frac{1}{8\pi G}\int_{\cC}\ve^{(0)}_{\cC} (\Omega\cL_\ell(f)-\cL_\ell(\Omega)f),\qquad P_Y'\hat{=}\frac{1}{8\pi G} \int_{\cC} Y^b(\pi_b+\varphi_b)\ve_{\cC}\,,
\ee
which, further assuming $\varphi_a=0$, is the starting point in \cite{Ciambelli:2026vxa} for the algebraic quantization.

In the next section, we will derive the Poisson and then the constrained Dirac brackets among the elementary fields on the hypersurface $\cN$. This analysis pertains to the bulk of the null hypersurface. Therefore, we restrict the bulk analysis to a null segment between caustics. Its endpoint cuts may be taken either as ordinary regulating cuts or as limiting cuts approaching caustic endpoints, provided the relevant boundary terms are well defined.

Nonetheless, it is interesting to discuss when fluxes vanish, because when that is the case the charges are canonically represented. Given \eqref{noeprime} and \eqref{fff}, the total flux of the system is
\beq\label{fluxes}
\cF^{\mathsf{can}}_\xi+Q_{\delta\lambda(\xi)}=\int_{\cC} \ell^c\ve_{c}\Big(f\Big(\frac{1}{2}\tau^{ab}\delta q_{ab}-\tau_a\delta \ell^a\Big)+\frac1{8\pi G}\, \delta\lambda(\xi)\Big)+\int_{\cC}Y^c\ve_c\Big(\frac{1}{2}\tau^{ab}\delta q_{ab}-\tau_a\delta \ell^a\Big).
\eeq
In this expression we used $\iota_{\xi}\ve_{\cN}=\xi^a\ve_a=f\ell^a\ve_a+Y^a\ve_a$ and we recall that $\lambda(\xi)=\ell(f)+Y^a\varphi_a$. Given the Ehresmann connection $k$, there is a family of adapted distributions ${\cal D}\subset T\cN$, which are sub-bundles of $T\cN$ spanned by vector fields that are in the kernel of $k$. If $k$ satisfies the Frobenius theorem, $\rd k\wedge k=0$, then these distributions are integrable, and one can find a codimension-$1$ sub-manifold of $\cN$, say $\cC$, such that ${\cal D}=T\cC$ is an integrable distribution. However, one notes that Frobenius theorem applies for the Ehresmann connection if and only if the vorticity $2$-form in eq.~\eqref{dk} vanishes, $\varpi=0$. Note that in the primed phase space the vorticity is a background structure that can be chosen at will.

Suppose we have vanishing vorticity, and we consider a family of adapted cuts $\cC$. By construction, the normal to the cut is $k_a$, such that $\ve_a=k_a \ve_{\cC}$, $Y^a\ve_a=0$, and $\ell^a\ve_a=\ve_{\cC}$. Therefore, the total flux \eqref{fluxes} at the cut becomes
\beq\label{totf}
\cF^{\mathsf{can}}_\xi+Q_{\delta\lambda(\xi)}=\frac1{8\pi G}\int_{\cC} \ve_{\cC}\Big(\frac{f}{2}(\sigma^{ab}-\mu q^{ab})\delta q_{ab}-f\pi_a \delta \ell^a+\delta(\ell(f)+Y^a\varphi_a)\Big),
\eeq
where we used $k_a\delta\ell^a=0$. Requiring $\delta \xi=\delta(f\ell^a+Y^a)=0$, and further assuming $\delta f=0$, we find $\delta Y^a=-f\delta \ell^a$. This is a new contribution with respect to \cite{Ciambelli:2023mir}, where we assumed $\delta\ell^a=0$. 

Then, noticing that in the primed phase space and in the absence of vorticity\footnote{In the presence of vorticity we have $ Y^a \delta \varphi_a =\delta\ell^a \varpi_{ab}Y^b$.} one has $Y^a\delta\varphi_a=0$, \eqref{totf} can be processed into 
\beq
\cF^{\mathsf{can}}_\xi+Q_{\delta\lambda(\xi)}=\frac1{8\pi G}\int_{\cC} \ve_{\cC}\Big(\frac{f}{2}(\sigma^{ab}-\mu q^{ab})\delta q_{ab}+\delta\ell^a(\overline D_a-\pi_a-\varphi_a) f\Big) .
\eeq
Using the unimodular decomposition of $q_{ab}$, we gather
\beq
\cF^{\mathsf{can}}_\xi+Q_{\delta\lambda(\xi)}=\frac1{8\pi G}\int_{\cC} \ve_{\cC}\Big(\frac{f\Omega}{2}\sigma^{ab}\delta\bq_{ab}-f\mu\delta \Omega+\delta \ell^a(\overline D_a-\pi_a- \varphi_a) f\Big).
\eeq
This total flux can be set to zero at the cut assuming that $f$ vanishes there. Alternatively, we can render the charges canonical by extending the phase space, as done in \cite{Donnelly:2016auv,Speranza:2017gxd,Ciambelli:2021nmv,Carrozza:2022xut}. While the latter is the proper way to account for the edge degrees of freedom, in this manuscript we confine our attention to the symplectic structure on the bulk of the null hypersurface, and study the constraints $C$ and $J_a$ on it. Therefore, we simply require that, either there are no cuts, or the total flux vanishes at the cuts. 

Assuming the total flux vanishes, the charge algebra can be computed and is given by
\beq
\{M'_f,M'_g\}&=&-M'_{[f,g]},\\
\{P'_Y,M'_f\}&=&-M'_{Y(f)},\\
\{P'_Y,P'_{Y'}\}&=& -P'_{[Y,Y']},
\eeq
where $[f,g]=f\ell(g)-g\ell(f)$ and $[Y,Y']$ is the horizontal Lie bracket, uniquely defined for $\varpi=0$.

\section{Kinematical Poisson Structure from Dirac Reduction}\label{s4}
In this section, we  derive the  Poisson bracket associated with the symplectic structure stemming from \eqref{thcan}.
This is a demanding task that involves considering all the nontrivial constraints among the phase-space variables. We start with \eqref{thcan}, that is,
\beq\label{tcans}
\Theta^{\mathsf{can}}=\int_{\cN} \ve_{\cN}^{(0)}\Big(\frac{1}{2}
\ttau^{ab}\delta q_{ab}
-\ttau_a\delta \ell^a\Big),
\eeq where we have defined the densitized fields 
\beq
\ttau^{ab}=\Omega\tau^{ab}\qquad \ttau_{a}=\Omega\tau_{a}.
\eeq
In the following, given that it is field independent, we will omit the bare measure $\ve_{\cN}^{(0)}$. Note that, using \eqref{tau}, defining $\tilde\sigma_{ab}=\Omega\sigma_{ab}$ and $\tilde\pi_a=\Omega \pi_a$, the symplectic potential can be elegantly recast as\footnote{Here, and in the remainder of the manuscript, we assume $8\pi G=1$. At the level of the brackets, it is simple to restore it, as any bracket among fields -- except for $\tau_a$ and $\tau_{ab}$ -- carries an $8\pi G$ in the right-hand side.}
\be \label{sympfinal}
\Theta^{\mathsf{can}} =
\int_{\cN}\left(\frac12 \psigma^{ab} \delta q_{ab} - \mu \delta \Omega - \ppi_a \delta \ell^a\right).
\ee 
Here, however, we will perform the symplectic reduction leading to \eqref{sympfinal} step by step, starting first from \eqref{tcans}, in which we first assume that $\tilde\tau_{ab}$ and $\tilde\tau_a$ are independent from $q_{ab}$ and $\ell^a$.

Therefore, the initial phase space consists of $4$ variables, $(\ttau^{ab},q_{ab},\ttau_a,\ell^a)$. The Poisson brackets are simply 
\beq\label{s0s0}
\PB{\ttau^{ab}(x_1)}{q_{cd}(x_2)}_0=2\delta^{ab}_{(cd)}\del{1}{2}\qquad \PB{\ttau_a(x_1)}{\ell^b(x_2)}_0=-\delta^b_a\del{1}{2},
\eeq
where $\delta^{ab}_{(cd)}=\frac12(\delta^{a}_{c}\delta^b_d+\delta^{a}_{d}\delta^b_c)$.
The subscript $0$ indicates that this bracket is the starting Poisson brackets which implements only the symmetries of $\ttau^{ab}$ and $q_{ab}$, and where the variables are unconstrained. We stress that this means that at this stage, $\ttau^{ab}$ and $\ttau_{a}$ are completely independent of $q_{ab}$ and $\ell^a$, and the only condition is that both $q_{ab}$ and $\ttau^{ab}$ are symmetric. Moreover, we introduced the notation $\del{i}{j}$, to emphasize that this is a densitized delta function, such that 
\beq\label{ddd}
\int_{2}\del{1}{2} f(x_2)=f(x_1),
\eeq
or, in other words $\del{1}{2}= \Omega(x_2)\delc{1}{2}$.
In the following we consistently use the tilde notation for densities and the un-tilde one for tensors. 

The procedure for building the bracket is composed of three steps, in which we impose constraints that transform this bare phase space into the gravitational phase space. These three steps are as follows:
\begin{itemize}
    \item Step 1: We first impose the constraints
    \be \ell^a k_a=1,\qquad \ell^a \tilde{\tau}_a= -\Omega\theta,
    \ee
    \item Step 2: We impose the orthogonality constraints
    \be \ell^a q_{ab}=0,\qquad \tilde{\tau}^{ab} k_b=0,
    \ee
    \item Step 3: We impose the shear constraint
    \be \Pi_{ab}{}^{cd}\left(
     \tau_{cd}- \frac12  \cL_{\ell} q_{cd}\right)=0.\label{shh}
    \ee
\end{itemize}
For the final constraint, we make use of the mixed-index tensor
\beq\label{PIPI}
\Pi_{ab}{}^{cd}=\frac12(q_{a}{}^{c}q_{b}{}^{d}+q_{a}{}^{d}q_{b}{}^{c}-q_{ab}q^{cd})
\eeq
which is the projector onto  symmetric and traceless rank-2 tensors, thus satisfying
\beq
\Pi_{ab}{}^{cd}\Pi_{cd}{}^{ef}=\Pi_{ab}{}^{ef}.
\eeq

The essential point is that, at every step, these constraints are second class. Indeed, at the first two steps the constraint matrices are algebraically invertible. At the third step, invertibility requires a choice of boundary conditions for the first-order operator along the null generators; we assume throughout that such an inverse exists on the selected function space. In other words, if we denote the constraints by $C_i$, the bracket 
\be 
\Upsilon_{ij}(x,y)=\{C_i(x),C_j(y)\}
\ee
is invertible. We can then define the Dirac Bracket at step $a+1$ to be 
\beq\label{Dirac}
\PB{O(x_1)}{O'(x_4)}_{a+1}=\PB{O(x_1)}{O'(x_4)}_{a}-\int_2\int_3 \PB{O(x_1)}{C_i(x_2)}_a (\Upsilon^{-1})^{ij}(x_2,x_3)\PB{C_j(x_3)}{O'(x_4)}_a.
\eeq
We adopt Einstein’s convention of implicit summation over the indices $i,j$. Furthermore, we use the subscript $a$ to indicate the constrained brackets after step $a$, on which we enforce the constraints $C_i$.

\subsection{First two Steps, Area Form, and Trace}

The construction of the reduced bracket $\{\cdot,\cdot\}_2$ obtained after the first and second step reduction is detailed in  appendices \ref{sec:First-step} and  \ref{sec:Second-step}. We report here the final result and discuss its implications in preparation of the third step.

After the second step reduction the metric $q_{ab}$ is degenerate due to the condition $\ell^a q_{ab}=0$. The brackets are also simpler when expressed in terms of $\tilde{\pi}_a =q_a{}^b\tilde{\tau}_b$ since the vertical component of $\tau_a$ is fixed.  Starting from \eqref{s0s0}, we find that the non-vanishing brackets among the elementary fields are given by 
\beq
\PB{\ppi_a(x_1)}{\ell^b(x_2)}_2&=&-q_a{}^b(x_1)\del{1}{2}\\
\PB{\ppi_a(x_1)}{\ppi_{b}(x_2)}_2&=&\big(\ppi_a(x_1)k_b(x_1)-k_a(x_1)\ppi_b(x_1)\big)\del{1}{2}\\
\PB{\ppi_a(x_1)}{q_{bc}(x_2)}_2
&=&(k_b(x_1)q_{ac}(x_1)+k_c(x_1)q_{ab}(x_1))\del{1}{2}\\
\PB{\ttau^{ab}(x_1)}{q_{cd}(x_2)}_2&=&(q_c{}^a(x_1)q_d{}^b(x_1)+q_d{}^a(x_1)q_c{}^b(x_1))\del{1}{2}.
\eeq
In order to move to step three, we need first to isolate the area form and the trace.

First, we extract the spin-$0$ data inside $q_{ab}$, using that 
\beq
q_{ab}=\Omega \bq_{ab}, \qquad q^{ab}=\frac{\bq^{ab}}{\Omega}
\eeq
with $\bq_{ab}$ unit determinant. Note that $\Omega$, being an area form, is nowhere vanishing, except at caustics, that are excluded from our analysis. In particular, we remark that $\bq_a{}^b=q_a{}^b=\delta_a{}^b - k_a \ell^b$. The bracket of $\Omega$ with the phase space variables can then be easily evaluated using that  
\beq
\PB{\Omega(x_1)}{F(x_2)}_2 =\frac12 
 \bq ^{ab}\PB{q_{ab}(x_1)}{F(x_2)}_2,
\eeq
which simply follows from
\be \label{Oq}
\delta \Omega =  \frac12 \Omega q^{ab }\delta q_{ab} =  \frac12\overline{q}^{ab} \delta q_{ab}.
\ee 
Using this we can evaluate that, at step two, $\Omega $ commutes with $(\Omega, \bq_{ab}, \ppi_a,\ell^a)$. Therefore, the only non trivial commutator involving $\Omega$ is with $\ttau^{ab}$ and it is given by 
\beq\label{tauO}
\PB{\ttau^{ab}(x_1)}{\Omega(x_2)}_2
=\bq^{ab}(x_2)\del{1}{2}.
\eeq
With this we can evaluate the brackets involving $\bq_{ab}$.

Like $q_{ab}$, $\bq_{ab}$ commutes with itself and with $\ell^a $ and $\ppi_a$.
The only modification involves $\ttau^{ab}$ and overall the non-zero brackets involving $\bq_{ab}$ are 
\beq
\PB{\ppi_a(x_1)}{\bq_{bc}(x_2)}_2&=&(k_b(x_1)\bq_{ac}(x_1)+k_c(x_1)\bq_{ab}(x_1))\del{1}{2},\\
\PB{\ttau^{ab}(x_1)}{\bq_{cd}(x_2)}_2 &=&\left(q_c{}^a(x_1)q_d{}^b(x_1)+q_d{}^a(x_1)q_c{}^b(x_1)-\bq^{ab}(x_1) \bq_{cd}(x_1)\right){\delc{1}{2}},\label{ttq}
\eeq
where we emphasize that in the last equation one has $ \delc{1}{2} = \del{1}{2} /\Omega(x_1)$. 

Before introducing the trace and traceless parts of $\tilde\tau^{ab}$, it is useful to lower its indices, and to convert it into $\tau_{ab}$ by extracting $\Omega$. Given that $q_{ab}$ commutes with itself, one has that the non-vanishing brackets are
\beq
\PB{\ppi_a(x_1)}{\ell^b(x_2)}_2&=&-q_a{}^b(x_1)\del{1}{2}\\
\PB{\ppi_a(x_1)}{\ppi_{b}(x_2)}_2&=&\big(\ppi_a(x_1)k_b(x_1)-k_a(x_1)\ppi_b(x_1)\big)\del{1}{2}\\
\PB{\ppi_a(x_1)}{\bq_{bc}(x_2)}_2&=&(k_b(x_1)\bq_{ac}(x_1)+k_c(x_1)\bq_{ab}(x_1))\del{1}{2},\\
\PB{\tau_{ab}(x_1)}{\bq_{cd}(x_2)}_2 &=&\left(\bq_{ac}(x_1)\bq_{bd}(x_1)+\bq_{ad}(x_1)\bq_{bc}(x_1)-\bq_{ab}(x_1) \bq_{cd}(x_1)\right)\del{1}{2}\\
\PB{\tau_{ab}(x_1)}{\Omega(x_2)}_2
&=&\bq_{ab}(x_2)\delc{1}{2}.\label{tO}
\eeq

Finally we can split $\tau_{ab}$ in its trace and traceless components
\be \label{traceless}
\tau_{ab}=-\mu q_{ab} +\hat{\tau}_{ab},\quad \text{with}\quad q^{ab}\hat\tau_{ab}=0.
\ee 
It is important at this stage to distinguish the traceless part of $\tau_{ab}$, here denoted $\hat\tau_{ab}$, from the shear $\sigma_{ab}$ appearing in \eqref{tau}. Indeed, at this point of the computation, we have not yet imposed \eqref{shh}, and thus $\hat\tau_{ab}$ is an independent symplectic datum. 

The symplectic fields are $(\Omega,\mu,\ell^a,\tilde\pi_a,\bq_{ab},\hat\tau_{ab})$. We have already computed all the brackets except those involving $\mu$ and $\hat\tau_{ab}$. They can be systematically extracted using \eqref{traceless} and taking the traceful and traceless parts of the brackets involving $\tilde\tau_{ab}$, while being careful of the factors of $\Omega$ involved. As an illustrative example, we extract the brackets $\PB{\Omega(x_1)}{{\mu}(x_2)}_2$ and $\PB{\Omega(x_1)}{{\hat\tau}_{ab}(x_2)}_2$ from \eqref{tO}. 

First, given that $\Omega$ commutes with itself, we have
\beq
\PB{\tau^{ab}(x_1)}{\Omega(x_2)}_2
=q^{ab}(x_2)\del{1}{2}.
\eeq
Then, contracting this with $q_{ab}(x_1)$ we get
\beq
q_{ab}(x_1)\PB{\tau^{ab}(x_1)}{\Omega(x_2)}_2=-2\PB{\mu(x_1)}{\Omega(x_2)}_2
=q_{ab}(x_1)q^{ab}(x_1)\del{1}{2}=2\del{1}{2},
\eeq
from which we deduce
\beq
\PB{\Omega(x_1)}{\mu(x_2)}_2
=\del{1}{2}.
\eeq
Then, we need to evaluate
\beq
\PB{\hat\tau^{ab}(x_1)}{\Omega(x_2)}_2
&=&q^{ab}(x_2)\del{1}{2}+\PB{\mu(x_1)q^{ab}(x_1)}{\Omega(x_2)}_2\\
&=&\mu(x_1)\PB{q^{ab}(x_1)}{\Omega(x_2)}_2.\label{comptin}
\eeq

Starting from
\beq
\PB{q_{a}{}^{b}(x_1)}{\Omega(x_2)}_2=-k_a(x_1)\PB{\ell^{b}(x_1)}{\Omega(x_2)}_2=0,
\eeq
one has
\beq
0=\PB{q_{a}{}^{b}(x_1)}{\Omega(x_2)}_2=q_{ac}(x_1)\PB{q^{bc}(x_1)}{\Omega(x_2)}_2,
\eeq
which implies that
\beq
\PB{q^{bc}(x_1)}{\Omega(x_2)}_2=\upsilon(x_1,x_2)\ell^b(x_1)\ell^c(x_1),
\eeq
for an unknown function $\upsilon(x_1,x_2)$. The latter is easily found to vanish, contracting with $k_b$ and using that it is a phase space constant (recall $\ell^a$ is assumed to be nowhere vanishing)
\beq
\upsilon(x_1,x_2)\ell^c(x_1)=k_b(x_1)\PB{q^{bc}(x_1)}{\Omega(x_2)}_2=\PB{k_b(x_1)q^{bc}(x_1)}{\Omega(x_2)}_2=0,
\eeq
such that eventually we have
\beq
\PB{q^{ab}(x_1)}{\Omega(x_2)}_2=0.
\eeq

Then, inserting this result back into \eqref{comptin}, we have derived
\beq
\PB{\hat\tau^{ab}(x_1)}{\Omega(x_2)}_2
=0\quad \Rightarrow\quad \PB{\hat\tau_{ab}(x_1)}{\Omega(x_2)}_2
=0\,.
\eeq

This strategy can be systematically employed to derive all the brackets involving $\hat\tau_{ab}$ and $\mu$. Therefore, after some computations one arrives at the final step two structure. The non-vanishing brackets are
\beq
\PB{\Omega(x_1)}{{\mu}(x_2)}_2&=&\del{1}{2}\\
\PB{\ell^a(x_1)}{\ppi_b(x_2)}{}_2&=&q_b{}^a(x_1)\del{1}{2}\\
\PB{\ppi_a(x_1)}{\ppi_{b}(x_2)}_2&=&\big(\ppi_a(x_1)k_b(x_1)-k_a(x_1)\ppi_b(x_1)\big)\del{1}{2}\\
\PB{\ppi_a(x_1)}{\bq_{bc}(x_2)}_2
&=&(k_b(x_1)\bq_{ac}(x_1)+k_c(x_1)\bq_{ab}(x_1))\del{1}{2}\label{pibq}\\
\PB{\ppi_a(x_1)}{\hat\tau_{bc}(x_2)}_2&=& (k_b(x_1)\hat\tau_{ac}(x_1)+k_c(x_1)\hat\tau_{ab}(x_1))\del{1}{2}\label{ss}\\
\PB{\bq_{ab}(x_1)}{\hat\tau_{cd}(x_2)}_2&=&\left(\bq_{ab}(x_1) \bq_{cd}(x_1)-\bq_{ac}(x_1)\bq_{bd}(x_1)-\bq_{ad}(x_1)\bq_{bc}(x_1)\right)\del{1}{2}\label{bqtt}\\
\PB{\hat{\tau}_{ab}(x_1)}{\hat{\tau}_{cd}(x_2)}_2&=&\left(\hat{\tau}_{ab}(x_1) \bq_{cd}(x_1)-\hat{\tau}_{cd}(x_1) \bq_{ab}(x_1)\right) \del{1}{2}
\eeq
Note that, importantly, \eqref{ss} is non-vanishing, even though $\PB{\ppi_a(x_1)}{\hat\tau^{bc}(x_2)}_2=0$. This is due to the fact that \eqref{pibq} is not zero, and thus lowering the indices of $\hat\tau^{ab}$ leads to \eqref{ss}. \\

We thus reached the end of the second step in our symplectic reduction. Our symplectic data are $(\Omega,\mu,\ell^a,\tilde\pi_a,\bq_{ab},\hat\tau_{ab})$ and, at this stage, the spin-0 $(\Omega,\mu$ and spin-1 $(\ell^a,\tilde\pi_a)$ sectors have been consistently reduced. The only remaining step is the one relating $\hat\tau_{ab}$ with the shear $\sigma_{ab}$, which completes our analysis for the spin-2 sector.

\subsection{The Third Step}\label{sec42}
 
In the third step, we must enforce the constraint \eqref{shh}. This is the hardest of the three steps. We report here only the main results, and delegate to Appendix \ref{AC} the various computational passages. 

The constraint  \eqref{shh} can be conveniently rewritten as
\beq
\Pi_{ab}{}^{cd} \tau_{cd}=\hat{\tau}_{ab}\stackrel{!}{=}\frac12 \Pi_{ab}{}^{cd} \cL_{\ell}q_{ab}=\sigma_{ab}.
\eeq
Given the unimodular decomposition, we have
\beq
\sigma_{ab}=\frac{\Omega}{2} \cL_{\ell}\bq_{ab}.
\eeq
Therefore, the constraint to impose is
\beq
\chi_{ab}=\hat{\tau}_{ab}-\frac{\Omega}{2} \cL_{\ell}\bq_{ab}.
\eeq
The bracket of the constraint can be computed and results in
\beq
\Upsilon_{abcd}(x_1,x_2)
&=&\PB{\chi_{ab}(x_1)}{\chi_{cd}(x_2)}_2\label{DDD}\\
&=&\Big(\hat{\tau}_{ab}(x_1)\bq_{cd}(x_1)-\bq_{ab}(x_1)\hat{\tau}_{cd}(x_1)\Big)\del{1}{2}
\nn\\&&+\Omega(x_1)\cL_{\ell(x_1)}(\overline\Pi_{abcd}(x_1)\del{1}{2})-\Omega(x_2)\cL_{\ell(x_2)}(\overline\Pi_{abcd}(x_1)\del{1}{2})\nn\,,
\eeq
where we used \eqref{bqtt}, which can be written as
\beq
\PB{\bq_{ab}(x_1)}{\hat\tau_{cd}(x_2)}_2=-2\overline\Pi_{abcd}(x_1)\del{1}{2}\qquad \overline\Pi_{abcd}=\Pi_{ab}{}^{ef}\bq_{ce}\bq_{df}\,.
\eeq

The inverse of the constraint bracket is explicitly derived in Appendix \ref{AC1}. It features the bilocal tensor $P_{ab}{}^{cd}$, which denotes a chosen inverse of the constraint kernel  satisfying $P_{abcd}(x_1,x_2)=-P_{cdab}(x_2,x_1)$. This tensor is subject to the differential equation
\beq\label{DiffP}
\delc{1}{2}\Pi_{ab}{}^{ef}(x_1)&=&\Big(\cL_{\ell(x_1)}-\frac{\theta(x_1)}{2}\Big)P_{ab}{}^{ef}(x_1,x_2)\\
&&-(\sigma_{ac}(x_1) q_{bd}(x_1)+ q_{ac}(x_1)\sigma_{bd}(x_1))P^{cdef}(x_1,x_2)\,.
\eeq
 The existence and uniqueness of this tensor depends on the choice of function space and boundary conditions for the first-order transport operator along the null generators. Throughout, we restrict to a null segment between caustics and assume boundary conditions for which this transport problem is well posed; possible zero/edge modes are understood to be fixed by these conditions, or equivalently projected out when defining the inverse. Different admissible boundary prescriptions may therefore correspond to different Green kernels, and we choose one satisfying the skew-symmetry required by the Poisson bracket.
 
To familiarize with this Green's function we can first write it explicitly choosing coordinates $x=(v,\theta^A)$ for which $\ell=\pa_v$ and 
\be 
\delc{1}{2} =
\frac{\delta(v_1-v_2) \delta^{(2)}(\theta_1,\theta_2)}{\sqrt{\Omega(x_1)\Omega(x_2)}}.
\ee
When the shear and expansion vanish, we have that $\Omega$ and $\Pi_{ab cd}$ are constant along the null time. Then, a solution of \eqref{DiffP} is given by 
 \be 
 P_{ab cd}(x_1,x_2) = \frac{H(v_1-v_2) \delta^{(2)}(\theta_1,\theta_2)}{\Omega(\theta_1)}  \Pi_{abcd}(\theta_1)
 \ee 
 where $H(v)=\frac12 \mathrm{sgn}(v)$ . 
 When the shear and expansion do not vanish we can obtain the general solution via a dressing involving the holonomy
 \be 
 U_a{}^b(x_1,x_2) = 
  \sqrt{
  \frac{\Omega(x_1)}{\Omega(x_2)}
  } \overleftarrow{\mathrm{P}}\!\!\exp\left(\int_{v_2}^{v_1} \sigma(v,\theta_1) \rd v \right)_a{}^b
 \ee 
 which is solution of the differential equation 
 $\pa_{v_1} U_a{}^b(x_1,x_2) = \left( \frac12 \theta(x_1) q_a{}^{a'}(x_1)+ \sigma_a{}^{a'}(x_1)\right) U_{a'}{}^b(x_1,x_2) $ with normalization $U_a{}^b(x_1,x_1)=1$.
 The tensor $P_{abcd}(x_1,x_2)$ is then given by 
 \be \label{OOPP}
 P_{ab cd}(x_1,x_2) = 
 \frac{H(v_1-v_2) \delta^{(2)}(\theta_1,\theta_2)}{\sqrt{\Omega(x_1)\Omega(x_2)}}  U_a{}^{a'}(x_1,x_2)U_b{}^{b'}(x_1,x_2)\Pi_{a'b'cd}(x_2) 
 \ee 
 The skew symmetry property follows from $U(x_2,x_1)= U^{-1}(x_1,x_2)$, $H(-v)=-H(v)$ and the covariance $U^{\otimes 4}(x_1,x_2) \Pi(x_2)=\Pi(x_1)$ valid when $x_1$ and $x_2$ are on the same null ray. In full generality, and thus when $\ell$ is arbitrary, the tensor $P_{abcd}$ is the generalization of \eqref{OOPP}, subject to \eqref{DiffP}.
This tensor is thus the covariant generalization of the propagator we found in \cite{Ciambelli:2023mir} using the Beltrami parametrization. As we are about to see, it plays the role of transporting symplectically the metric between points, that is, to propagate the degrees of freedom on $\cN$.

Before applying Dirac's formula \eqref{Dirac}, we need to evaluate the brackets between the symplectic fields and the constraint $\chi_{ab}$. This can be easily carried out for all of the fields, except for $\PB{\ppi_{a}(x_1)}{\chi_{cd}(x_2)}_2$, which is a more challenging computation we perform in Appendix \ref{AC2}. The final outcome is ($\overline{\pa}_{a}^2=q_a{}^b(x_2)\frac{\pa}{\pa x^b_2}$)
\beq
\PB{\Omega(x_1)}{\chi_{cd}(x_2)}_2&=&0\\
\PB{\pmu(x_1)}{\chi_{cd}(x_2)}_2&=&
\delc{1}{ 2} \sigma_{cd}(x_2)\\
\PB{\ell^a(x_1)}{\chi_{cd}(x_2)}_2&=&0\\
\PB{\ppi_a(x_1)}{\chi_{bc}(x_2)}_2
&=&\Big(k_a(x_2)\sigma_{bc}(x_2)-\frac{\Omega(x_2)}{2} q_b{}^d(x_2)q_c{}^e(x_2) \pa^{2}_a \bq_{de}(x_2)+\frac12\big(q_{bc}(x_2)(\overline \pa^{2}_a+\varphi_a(x_2))\nn\\
&&-q_{ab}(x_2)(\overline\pa^{2}_c +\varphi_c(x_2))-q_{ac}(x_2)(\overline\pa^{2}_b +\varphi_b(x_2))\big)\Big)\del{1}{2}\\
\PB{\bq_{ab}(x_1)}{\chi_{cd}(x_2)}_2
&=&-2\overline{\Pi}_{abcd}(x_1)\del{ 1}{ 2}\\
\PB{\hat\tau_{ab}(x_1)}{\chi_{cd}(x_2)}_2
&=& (\sigma_{ab}(x_1)q_{cd}(x_1)-q_{ab}(x_1)\sigma_{cd}(x_1))\delc{1}{2}\\
&& - \Omega(x_2)\cL_{\ell(x_2)}(\overline\Pi_{abcd}(x_2) \del{ 1}{ 2})
\eeq

Armed with this, we can eventually compute the brackets at step three. This is a long and highly non-trivial computation, which we explain in Appendix \ref{AC3}. The brackets now involve only the fields $(\Omega,\mu,\ell^a,\tilde\pi_a,\bq_{ab})$, since at this step $\hat\tau_{ab}=\sigma_{ab}$ is no longer an independent datum. This means that, while one can certainly evaluate brackets involving $\sigma_{ab}$ using Dirac's formula, this is not needed: it is by consistency that the latter can be derived from brackets involving $\bq_{ab}$ by taking its Lie derivative along $\ell^a$ and multiplying by $\Omega$. We have checked explicitly that this is the case, subject to the differential equation \eqref{DiffP}. 

It turns out to be more elegant and instructive to write the brackets involving $q_{ab}$ instead
of $\bq_{ab}$, such that the symplectic data are $(\Omega,\mu,\ell^a,\tilde\pi_a,q_{ab})$. The final non-vanishing brackets are therefore
\beq
\PB{\Omega(x_1)}{{\mu}(x_2)}_3&=&\del{1}{2}\label{Omu}\\
\PB{\mu(x_1)}{\mu(x_2)}_3&=&\frac12 \sigma_{ab}(x_1)P^{abcd}(x_1,x_2)\sigma_{cd}(x_2)\label{mm}\\
\PB{\mu(x_1)}{\tilde\pi_{a}(x_2)}_3&=&-\frac14\sigma_{cd}(x_1) P^{cdef}(x_1,x_2)\Omega(x_2)\cL_{\overline\pa^2_a}q_{ef}(x_2)+\frac12\paleft_e^2\left(\sigma_{cd}(x_1) P^{cde}{}_{a}(x_1,x_2)\Omega(x_2)\right)\nn\\
\PB{\mu(x_1)}{q_{ab}(x_2)}_3&=&-q_{ab}(x_2)\delc{1}{2}-\sigma_{cd}(x_1) P^{cd}{}_{ab}(x_1,x_2)\\
\PB{\ell^a(x_1)}{\ppi_b(x_2)}{}_3&=&q_b{}^a(x_1)\del{1}{2}\label{pl}\\
\PB{\tilde\pi_{a}(x_1)}{\tilde\pi_{b}(x_2)}_3&=&(\ppi_a(x_1) k_b(x_1)-\ppi_b(x_1)k_a(x_1))\del{1}{2}\label{pp}\\
&&+\frac18 \cL_{\overline\pa_a^1}q_{cd}(x_1)P^{cdef}(\tilde x_1,\tilde x_2)\cL_{\overline\pa_b^2}q_{ef}(x_2)+\frac12\paleft_d^1 \paleft_e^2 P_a{}^{de}{}_b(\tilde x_1,\tilde x_2)\nn\\
&&-\frac14(\paleft_d^1 P_a{}^{def}(\tilde x_1,\tilde x_2))\cL_{\overline\pa_b^2}q_{ef}(x_2)-\frac14(\paleft_e^2P^{cde}{}_b(\tilde x_1,\tilde x_2))\cL_{\overline\pa_a^1}q_{cd}(x_1)\nn\\
\PB{\ppi_a(x_1)}{q_{bc}(x_2)}_3&=&(k_b(x_1)q_{ac}(x_1)+k_c(x_1)q_{ab}(x_1))\del{1}{2}+\frac{\Omega(x_1)}{2}\cL_{\overline\pa^1_a}q_{de}(x_1) P^{de}{}_{bc}(x_1,x_2)\nn\\
&&-\paleft^1_d\left(\Omega(x_1) P^{d}{}_{abc}(x_1,x_2)\right)\label{pqpq}\\
\PB{q_{ab}(x_1)}{q_{cd}(x_2)}_3&=&2 P_{abcd}(x_1,x_2)\,,\label{qq}
\eeq
where we defined $\paleft^1_a=\pa^1_a-\varphi_a(x_1)$, $\cL_{\overline\pa_a}q_{cd}=\pa_aq_{cd}-k_a\cL_{\ell}q_{cd}$, and we recall $P_{abcd}(\tilde x_1,\tilde x_2)=\Omega(x_1)P_{abcd}(x_1,x_2)\Omega(x_2)$.

This is the main result of the paper, and thus deserves various comments on its mathematical structure and physical interpretation. 

First, we reiterate that all the brackets involving $\sigma_{ab}$ can now be deduced from the brackets above. In fact, with this last step of the computation we have imposed all the kinematical constraints. This is the most general phase space of gravity on null hypersurfaces, prior to imposing the Raychaudhuri and Damour constraints. 

Secondly, this generalizes the analysis performed in \cite{Ciambelli:2023mir} to include the spin-1 data.\footnote{Albeit with different notation and conventions, the careful analysis performed in \cite{Adami:2023wbe} can be recovered from our analysis (prior to imposing the Raychaudhuri and Damour constraints) by gauge fixing $\delta\ell^a=\delta^a_A \delta U^A$, where $x^A$ is the horizontal coordinate.} This was not a straightforward task, as the brackets involving $\ppi_a$ turn out to be quite complicated. As we saw in \cite{Ciambelli:2023mir}, the symplectic partner of $\Omega$, $\mu$, acts via Poisson brackets by inserting the shear, i.e., the temporal Lie derivative of the metric. This is readily encapsulated in \eqref{mm}. Here, we have found that the field $\ppi_a$ plays an analogue role by inserting the horizontal Lie derivative of the metric. Indeed, let us rewrite \eqref{mm} with the shear explicit and the bilocal term inside \eqref{pp} next to each other,
\beq
&\PB{\mu(x_1)}{\mu(x_2)}_3=\frac18 \cL_{\ell(x_1)}\bq_{ab}(x_1)P^{abcd}(\tilde x_1,\tilde x_2)\cL_{\ell(x_2)}\bq_{cd}(x_2)&\\
&
\PB{\tilde\pi_{a}(x_1)}{\tilde\pi_{b}(x_2)}_3\supset\frac18 \cL_{\overline\pa_a^1}q_{cd}(x_1)P^{cdef}(\tilde x_1,\tilde x_2)\cL_{\overline\pa_b^2}q_{ef}(x_2)\,.&
\eeq
This clearly demonstrates how the brackets of these two fields involve the evolution in time and space of the metric, respectively. 

Third, we also see why we named $P_{abcd}$ the propagator. Indeed, from \eqref{qq}, one has that $P_{abcd}$ relates the degenerate metric at different points. Then, all the other brackets have either a local term $\del{1}{2}$, or this propagator -- and its derivatives. Actually, one can push this interpretation even further by introducing a diagrammatic way to understand these brackets: each $\mu$ inserts a $\sigma_{ab}$, each $\ppi_a$ inserts a $\cL_{\overline\pa_a}q_{bc}$, and each index contraction and bilocal expression is performed using $P_{abcd}$. It would be interesting to push this interpretation even further.

Lastly, these brackets should be taken as the starting point for the kinematic quantization. This was the input in the analysis performed in \cite{Ciambelli:2024swv}, where only the spin-0 and later the perturbative spin-2 sectors were considered. Here, we have in principle the general phase space. As we will explain in the conclusions, it would be rewarding to study this general setup and discuss the fate of the anomaly found in \cite{Ciambelli:2024swv}, once we include the Damour constraint, and thus we move from a single light ray to a collection of them. 

\section{Alternative Derivation via Phase Space Variations}\label{s5}

The brackets derived in the previous section can be reconstructed directly from the reduced symplectic form by identifying the Hamiltonian vector fields generated by a complete set of smeared observables. This requires recasting the symplectic two form in a way that clearly isolates each spin sector of the theory. 

We denote the variation of the degenerate metric $q_{ab}$ and its various contractions as 
\be 
\Delta_{ab}= q_a{}^cq_b{}^d \delta q_{cd}, \qquad 
\hat\Delta_{ab} =\Pi_{ab}{}^{cd}\delta q_{cd},
\qquad 
\Delta = q^{ab} \delta q_{ab}=2\frac{\delta\Omega}{\Omega}, \qquad 
\delta \ell_a := q_{ab}\delta \ell^b.
\ee
We remark that these tensors are all horizontal so we can raise and lower the indices with $q_{ab}$.
Then, we have
\begin{align} 
\delta q_{ab} &= \hat{\Delta}_{ab} + q_{ab}\frac{\Delta}{2}  - k_a \delta \ell_b - k_b \delta \ell_a\\
\delta q^{ab} &= -\hat{\Delta}^{ab} - q^{ab} \frac{\Delta}{2}= - \Delta^{ab} .
\end{align} 
Denoting $\Pi_{\delta\ell bcd}=\delta\ell^a\Pi_{abcd}$, a useful identity for the variation of the projector is
\beq \label{varPi}
\delta \Pi_{abcd}&=& \frac12[\hat{\Delta}_{ab}q_{cd} + \hat{\Delta}_{cd} q_{ab}] + \Pi_{abcd} \Delta - 
 (k_{a} \Pi_{\delta\ell bcd} + k_b \Pi_{a\delta\ell cd} + k_c \Pi_{ab\delta\ell d}+ k_d \Pi_{abc\delta\ell}).\\
 \delta \Pi^{abcd} &=& - \left(\frac12 \hat{\Delta}^{ab} q^{cd}  + \frac12 q^{ab}\hat{\Delta}^{cd} +\Delta \Pi^{abcd} \right).\label{varPi2}
\eeq 
The validity of these formulas can be checked by contracting both sides with $(q^{ab}, q^{bc}, \ell^a)$. A detailed proof is given in appendix \ref{sec:variation}.

Next, introducing the compact notation $(\Delta \!\cdot\! \theta)^{ab}= \Delta^a{}_{c} \theta^{cb}$ and $\Delta\!:\!\sigma= \Delta^{ab}\sigma_{ab}$, we evaluate
\beq
  \delta \theta^{ab}   &=& \frac12 q^{aa'}q^{bb'} \cL_{\ell} \Delta_{a'b'} -  2(\Delta \cdot \theta)^{(ab)} + (\oD + \varphi)^{(a}\delta \ell^{b)}, \\
   \delta \theta
 &=& \frac12 \cL_{\ell} \Delta  +(\oD_a+\varphi_a) \delta \ell^a \label{dtheta}\\
       \delta \sigma^{ab}
       &=& \delta \Pi^{abcd} \theta_{cd} + \Pi^{abcd} \delta \theta_{cd} \\
    &=&   \Pi^{abcd}\left(\frac12(\cL_{\ell} -\theta)\hat\Delta_{cd}+ (\oD_c+\varphi_c) \delta \ell_d \right) -\frac12 \Delta \sigma^{ab} -\frac12 (\Delta \!:\!\sigma) q^{ab}.
 \eeq
To derive these expressions, we used  $\cL_{\ell}k_a=-\varphi_a$ and the identity ($\oD^a Y^b=q^{ac}q_d{}^b \oD_cY^d$)
\be 
 q_a{}^cq_b{}^d \cL_{Y} q_{cd}= \oD_a Y_b +\oD_b Y_a,
\ee
which holds when $Y$ is an horizontal vector.
Another useful identity is  
\be 
(\Delta \!\cdot\! \theta)^{(ab)} = \frac12\left(
\Delta^{ab}\theta +  \Delta \sigma^{ab} + (\hat\Delta\!:\!\sigma) q^{ab}\right).
\ee 

From the variation of the shear we conclude that 
 \begin{align}
    \delta \psigma^{ab}
    &= -\frac{\Omega}2 
    (\hat\Delta\!:\!\sigma) q^{ab}  
    + \Omega \Pi^{abcd}\left(\frac12(\cL_{\ell}-\theta) \hat{\Delta}_{cd} +  (\oD_c +\varphi_c)\delta \ell_d\right).\label{dsigma}
\end{align}
Therefore, using that  $\Omega \Delta =2 \delta \Omega=\Omega q^{ab}\delta q_{ab}$, $\Pi^{abcd}\delta q_{ab}=\Pi^{abcd}\hat\Delta_{ab}$, and $\Pi_{abcd}=\Pi_{cdab}$, we have that 
\begin{align}
&\,\delta \psigma^{ab} \wedge \delta q_{ab} = \frac12 \Omega \Delta   \wedge (\hat\Delta\!:\! \sigma) +\frac12 \Omega \Pi^{abcd} \cL_{\ell} \hat{\Delta}_{ab} \wedge \hat{\Delta}_{cd} 
    + \Omega \Pi^{abcd} (\oD_a+\varphi_a)\delta \ell_b \wedge \hat\Delta_{cd}.
\end{align}

We can now evaluate the integrated symplectic form. Starting from \eqref{sympfinal}, and assuming that all boundary terms produced by integrations by parts -- both on the cuts and at the endpoint cuts of the null segment -- vanish, or are treated by boundary conditions compatible with the chosen Green kernel, we obtain
\beq 
\Omega^{\mathsf{can}}&=&
\int_{\cN} \delta\Omega    \wedge \left(\delta \mu +\frac12 (\hat\Delta\!:\! \sigma)
\right) +\frac14  \Pi^{abcd} \cL_{\ell} (\sqrt{\Omega}\hat{\Delta}_{ab}) \wedge (\sqrt{\Omega}\hat{\Delta}_{cd})\\
&&
    +\delta\ell^a \wedge \left(\delta\ppi_a - \frac{\Omega}{2}    (\oD^b -\varphi^b)\hat\Delta_{ab} \right)\nonumber
\eeq 
This expression cleanly identifies the spin $0$, spin $1$ and spin $2$ momenta. We see that the spin $2$ variables affect the spin $0$ and spin $1$ momenta through the shifts 
$\delta \mu \to \delta \mu +\frac12 
(\hat{\Delta}\!:\!\sigma)$ and $\delta \ppi_a\to \delta \ppi_a - \frac{\Omega}{2}[(\oD-\varphi)\!\cdot\!\hat\Delta]_a$. Note that, importantly, given this processed symplectic two-form, the associated brackets are the step-3 reduced brackets of the previous section, since all the constraints of the previous steps are directly imposed.

\subsection{Spin-2 Hamiltonian Generators}

Now that we have built the symplectic two-form we can construct the charges associated respectively with the spin $2$, spin $1$ and spin $0$ variables.
We start with the spin $2$ charges.

To construct them we make use of the bilocal propagator $P_{ab}{}^{cd}(x,y)$ we introduced in the previous section. We recall its properties are $\Pi_{ab}{}^{a'b'}(x)P_{a'b'}{}^{cd}(x,y)=P_{ab}{}^{cd}(x,y) = P_{ab}{}^{c'd'}(x,y)\Pi_{c'd'}{}^{cd}$ and $P_{abcd}(x,y)=-P_{cd ab}(y,x)$. It can be used to define an endomorphism acting on  symmetric traceless quantities
\be 
{P}_{ab}[{S}](x_1):=\int_2 P_{ab}{}^{cd}(x_1,x_2) {\Omega}(x_2) {S}_{cd}(x_2)\,,
\ee 
where the factor of $\Omega$ ensures the proper normalization of the measure, compatible with \eqref{ddd}.
This map is such that $\Pi_{ab}{}^{cd} {P}_{cd}[{S}]= {P}_{ab}[{S}]$, and it is chosen to be  Green's function for the Lie derivative acting on traceless symmetric tensors
\be \label{pop}
\left(\cL_{\ell}-\frac12 \theta\right) {P}_{ab}[{S}] -   (  \sigma\!:\!{P}[{S}]) q_{ab} = \hat{S}_{ab}\,,
\ee 
where $\hat{S}_{ab}=\Pi_{ab}{}^{cd}S_{cd}$. 
From this we deduce the local expression
\beq\label{Pdiff}
\delc{1}{2}\Pi_{ab}{}^{ef}(x_1)&=&\Big(\cL_{\ell(x_1)}-\frac{\theta(x_1)}{2}\Big)P_{ab}{}^{ef}(x_1,x_2)\\
&&-(\sigma_{ac}(x_1) q_{bd}(x_1)+ q_{ac}(x_1)\sigma_{bd}(x_1))P^{cdef}(x_1,x_2)\,,
\eeq
which is nothing but \eqref{DiffP}. Note that, while this last expression is valid in any $d$, \eqref{pop} is true only when $\cN$ is $3$-dimensional.

Given this map we can define the spin $2$ action on the Carrollian geometry to be  
\begin{align}
\fL'_S q_{ab} &= P_{ab}[{S}],\quad \fL'_S  \ell^a=0,\quad 
\fL'_S \Omega =0.
\end{align}
The action on the momenta is then given by 
\beq
\fL'_S\ppi_a &=&\frac12  \Omega (\oD^b-\varphi^b) P_{ab}[S] \\
\fL'_S \mu &=& -\frac12 (\sigma\!:\! P[S]).
\eeq
The action on the  shear can be deduced from (\ref{dtheta},\ref{dsigma}). They are given by 
\beq
 \fL'_S\psigma^{ab}&=&\frac{\Omega}{2} \left(\Pi^{abcd}(\cL_{\ell}-\theta) P[S]_{cd}  - (\sigma\!:\! P[S]) q^{ab}\right)\\
 &=&\frac{\Omega}{2}\left( \hat{S}^{ab} - \frac{\theta}{2} P^{ab}[S]- (\sigma\!:\! P[S]) q^{ab} \right) \\
\fL'_S \theta &=&  0.
\eeq

This transformation is canonical. We find, after integration by parts\footnote{We use that, for any horizontal symmetric tensor $X_{ab}$,
\be
(\cL_{\ell} \Pi^{abcd}) X_{cd} =-( (\sigma\!:\!\hat{X}) q^{ab}+ X \sigma^{ab}+2\theta\hat X^{ab}),
\ee 
so that 
$\hat{\Delta}_{ab}(\cL_{\ell} \Pi^{abcd})P_{ab}[S] =-2\theta\hat{\Delta}_{ab} P^{ab}[S]$.} 
and using \eqref{pop} that 
\beq
I'_{S}\Omega^{\mathsf{can}}
&=& \frac1{2}
\int_{\cN}  \Omega\hat\Delta_{ab}(\Pi^{abcd}\cL_{\ell}P_{cd}[S]+\frac{\theta}{2}P^{ab}[S]+\frac12 (\cL_{\ell}\Pi^{abcd})P_{cd}[S])  \\
&=&\frac1{2}
\int_{\cN}  \Omega\hat\Delta^{ab}(\Pi_{ab}{}^{cd}\cL_{\ell}P_{cd}[S]-\frac{\theta}{2}P_{ab}[S])\\
&=&  \frac1{2}
\int_{\cN}  S_{ab}  \Omega \Pi^{abcd}\delta q_{cd}\\
&=& - \delta \left(\frac1{2}
\int_{\cN}  S_{ab} \overline{q}^{ab}\right),
\eeq
where we use in the last equality that $\delta \overline{q}^{ab}= - \Omega \Pi^{abcd}\delta q_{cd}$ and  that $\delta S_{ab}=0$.
This shows that the charge 
\be 
Q_S:=
\frac1{2}
\int_{\cN} S_{ab} {\bq}^{ab}\,,
\ee 
is the Hamiltonian generator of the transformation $\fL'_S$. 
That is, for any phase-space observable $\mathcal O(x_1)$, the Hamiltonian
action generated by $Q_S$ is\footnote{As we already mentioned, these brackets must coincide with the step-3 brackets of the previous section. We omit the label $3$ in this section.} (recall $\bq^{ab}=\Omega q^{ab}$)
\beq
\mathcal L'_S\mathcal O(x_2)
&=&
\{Q_S,\mathcal O(x_2)\}
\\
&=&
\frac12\int_1
S_{ab}(x_1)
\bigl\{
\overline q^{ab}(x_1),\mathcal O(x_2)
\bigr\}
\\
&=&
-\frac12\int_1
\Omega(x_1)\hat S^{ab}(x_1)
\bigl\{
q_{ab}(x_1),\mathcal O(x_2)
\bigr\}
\\
&=&-\frac12\int_{1}
\Big[
\Omega(x_1)S^{ab}(x_1)
\bigl\{
q_{ab}(x_1),\mathcal O(x_2)
\bigr\}
-
S(x_1)
\bigl\{
\Omega(x_1),\mathcal O(x_2)
\bigr\}
\Big].
\label{eq:spin-two-generator}
\eeq

\subsection{Spin-0 Hamiltonian Generators}

We can now focus on the spin $0$ sector. There are two actions: the trivial one parametrized by a scalar field $\alpha$ leaves the Carrollian geometry invariant 
\be 
\fL'_\alpha q_{ab}=0,\quad \fL'_\alpha \ell^a=0
\ee 
while its action on the momenta is simply 
\be \label{Lam}
\fL'_\alpha \mu = \alpha, \quad \fL'_\alpha \ppi_a =0, \quad \fL'_\alpha \psigma^{ab}=0.
\ee 
This action is canonical with charge given by the local area element $\Omega_\alpha:= 
 \int_{\cN} \alpha \Omega$:
\be \label{alal}
I'_\alpha \Omega^{\mathsf{can}} = - \delta \Omega_\alpha, \qquad \{\Omega_\alpha,\cO\}=\fL'_\alpha \cO.
\ee 

The second spin $0$ action, labeled by the scalar field\footnote{$\alpha$ and $\beta$ are assumed to be field independent so that $\delta\alpha=\delta \beta=0$.} $\beta$, acts non-trivially on the  Carrollian geometry. It is  given by   
\begin{align}
\fL'_\beta q_{ab} &= \beta q_{ab} -   P_{ab}[ \beta \sigma],\quad \fL'_\beta   \ell^a=0,\quad 
\fL'_\beta \Omega =\beta \Omega  .
\end{align}
In the first equation, the first term changes the area element, whereas the nonlocal $P_{ab}[ \beta \sigma]$ term is the compensating traceless deformation required for the transformation to be Hamiltonian with respect to the full spin-two-coupled symplectic form.

The action on the expansion and momenta is then given by 
\beq
\fL'_\beta\ppi_a & =&-  \frac12\Omega (\overline{D}^b-\varphi^b) P_{ab}[\beta \sigma] \\
\fL'_\beta \mu &=&-\beta \mu + \frac12 (\sigma\!:\! P[\beta \sigma]) \\
\fL'_\beta \theta &=&\cL_\ell \beta \\
\fL'_\beta\psigma^{ab}
 &=& \frac{\Omega}{2}  \left(  (\sigma\!:\! P[\beta \sigma]) q^{ab} -\beta\sigma^{ab}+\frac{\theta}{2} P^{ab}[\beta \sigma]\right).
\eeq

Using the transformations above, the spin-zero and spin-one components
of the contraction of the symplectic form reduce to
\begin{align}
I'_\beta\Omega^{\mathsf{can}}_{(0+1)}
=
\int_{\mathcal N}
\left[
\delta(\beta\Omega\mu)
+\frac{\beta\Omega}{2}
(\sigma\!:\!\hat\Delta)
\right].
\end{align}
The remaining spin-two contribution is
\beq
I'_\beta\Omega^{\mathsf{can}}_{(2)}
&=&
-\frac{1}{2}
\int_{\mathcal N}
\Omega\,\hat\Delta^{ab}
\left[
\Pi_{ab}{}^{cd}\mathcal L_\ell
P_{cd}[\beta\sigma]
-\frac{\theta}{2}P_{ab}[\beta\sigma]
\right]\\
&=&
-\frac{1}{2}
\int_{\mathcal N}
\beta\Omega\,(\sigma\!:\!\hat\Delta),
\eeq
where in the last equality we used the Green equation \eqref{pop}.
The spin-two terms cancel, and hence we find that the action is canonical with charge $m_\beta := \int_{\mathcal N}\beta\Omega\mu$
\begin{align}
I'_\beta\Omega^{\mathsf{can}}
=
\delta
\left(
\int_{\mathcal N}\beta\Omega\mu
\right)=-\delta m_\beta.
\end{align}
This means that 
\be 
\{m_\beta,\cO(x_2)\}=-\int_{1} \{\Omega(x_1) \mu (x_1),\cO(x_2)\} \beta(x_1) = \fL'_\beta \cO(x_2).
\ee

\subsection{Spin-1 Hamiltonian Generators}
We can finally  focus on the spin $1$ sector. As per the spin $0$ one, there are two actions: the trivial one parametrized by a field independent one-form field $A_a$ leaves the Carrollian geometry invariant 
\be 
\fL'_A q_{ab}=0,\quad \fL'_A \ell^a=0,
\ee 
while its action on the momenta is simply 
\be \label{LAMB}
\fL'_A \mu = 0, \quad \fL'_A \ppi_a =q_a{}^b A_b, \quad \fL'_A \psigma^{ab}=0.
\ee 
This action is canonical with charges given by $L_A:= 
\int_{\cN} A_a \ell^a$:
\be \label{LA}
I'_A \Omega^{\mathsf{can}} = - \delta L_A, \qquad \{L_A,\cO\}=\fL'_A\cO.
\ee 
The second spin $1$ action is labeled by a vector field  $B^a$ and acts non-trivially on the  Carrollian geometry, 
\begin{align}\label{LBS}
\fL'_B q_{ab} &= - P_{ab}[(\oD+\varphi)B ]-k_a B_b -k_b B_a,\quad \fL'_B  \ell^a=B^a,\quad 
\fL'_B \Omega =0,
\end{align}
where $P_{ab}$ acts upon the symmetric tensor $[(\oD+\varphi)B]_{ab}:= (\oD_{(a}+\varphi_{(a})B_{b)}$. Note that $\fL'_B\ell^a=B^a$ implies that $B^ak_a=0$ and thus $B^a$ is horizontal.
The action on the momenta is then given by 
\beq
 \fL'_B\psigma^{ab}
 &=&\frac{\Omega}{2}\left( \Pi^{abcd}(\oD_c+\varphi_c)B_d + \frac{\theta}{2} P^{ab}[(\oD+\varphi)B]+(\sigma\!:\!P[(\oD+\varphi)B]) q^{ab}\right), \\
\fL'_B\ppi_a & =& -\frac{\Omega}{2} (\oD^b-\varphi^b) P_{ab}[(\oD+\varphi)B] - k_a \ppi_b B^b, \label{LBp}\\
\fL'_B \mu &=&  \frac12(\sigma\!:\! P[(\oD+\varphi)B])\\
\fL'_B \theta &=&  (\oD_a+\varphi_a)B^a.
\eeq
This transformation is canonical. Again integrating by parts, recalling 
$\Omega \Delta =2 \delta \Omega$ and $\Pi^{abcd}\Delta_{cd}= \hat{\Delta}^{ab}$, the spin-two contribution
cancels the non-exact part of the spin-one contribution, and one finds
\begin{equation}\label{pbpb}
I'_B\Omega^{\mathsf{can}}
=
\delta
\left(
\int_{\mathcal N}B^a\tilde\pi_a
\right)=-\delta p_B.
\end{equation}
This means that 
\be 
\{p_B,\cO(x_2)\}=-\int_{1} \{\ppi_a(x_1),\cO(x_2)\} B^a(x_1) = \fL'_B \cO(x_2).
\ee 

This concludes the computation of the canonical Hamiltonians for each spin sector. We can now confirm the compatibility of the brackets found above with these symplectic actions.

\subsection{Poisson Brackets from the Hamiltonian Generators}

The charges $Q_S,\Omega_\alpha, m_\beta, L_A, p_B$ are the generating functionals for the respective symplectic actions
$(\fL'_S,\fL'_\alpha,\fL'_\beta,\fL'_A,\fL'_B)$. Having constructed the Hamiltonian generators directly from the
symplectic two-form, we can now reconstruct the elementary 
brackets. For each field-independent smearing parameter, the Hamiltonian
identity
\begin{equation}
    \fL'_X\mathcal O=\{Q_X,\mathcal O\}
\end{equation}
holds for an arbitrary phase-space observable $\mathcal O$. Since the
smearing parameters $S_{ab}$, $\alpha$, $\beta$, $A_a$, and $B^a$ are
arbitrary, these identities can be unsmeared to determine the
distributional kernels of the elementary brackets. The resulting
expressions agree with the Dirac brackets obtained independently in
Section \ref{sec42}, providing an important independent verification of these brackets.

Let us begin by explicitly showing the simplest example, to explain the logic. Consider \eqref{alal}, and specify $\cO=\mu$, then 
\beq
\{\Omega_\alpha,\mu(x_2)\}=\int_1 \alpha(x_1)\{\Omega(x_1),\mu(x_2)\}=\fL'_\alpha\mu(x_2)=\alpha(x_2)\,,
\eeq
where we used \eqref{Lam} in the last equality. For this equation to hold, we must enforce
\beq
\{\Omega(x_1),\mu(x_2)\}=\del{1}{2}\,,
\eeq
which is exactly what we found in \eqref{Omu}. The field space actions for $\Omega_\alpha$ also directly gives that  $\Omega$ commutes with all the variables except $\mu$.

Very similarly, and equally straightforward, is the confirmation of \eqref{pl}, that is,
\begin{align} 
\{\ell^a(x_1), \ppi_{b}(x_2)\}&=q_a{}^b \del{1}{2},
\end{align} 
from \eqref{LA} and \eqref{LAMB}, and the fact that $\ell^a$ commutes with all observables except $\ppi_a$.

From the action of $Q_S$ and $m_\beta$ we can deduce the commutation relations of $q_{ab}$ with all the fields
\beq
\{q_{ab}(x_1),q_{cd}(x_2)\}
&=&2 P_{abcd}(x_1,x_2)\\
\{\mu(x_1),  q_{cd}(x_2)\}
&=& -\left(\sigma^{ab}(x_1) P_{abcd}(x_1,x_2)+ q_{cd}(x_2) \delc{1}{2}\right).
\eeq
 
For the spin-$1$ sector, the generator $p_B$ in \eqref{pbpb} can be used to compute the brackets.
To show the equivalence with the previous computation we use that $\frac12 \cL_{\overline \pa_a} q_{bc}P^{bc}{}_{de}= q_{a}{}^b \Gamma_{bf}^{c}P^{f}{}_{cde}$ to express the brackets (\ref{pp}, \ref{pqpq}) in a more compact form involving covariant derivative 
\beq
    \PB{\tilde\pi_{a}(x_1)}{\tilde\pi_{b}(x_2)}&=&\frac12 
 \Omega(x_1)\Omega(x_2) (\overline{D}_1^c-\varphi^c(x_1))(\overline{D}_2^d-\varphi^d(x_2)) P_{ac bd}(x_1,x_2)\\
 &&
+
\big(\ppi_a(x_1)k_b(x_1)-k_a(x_1)\ppi_b(x_1)\big)\del{1}{2} \\
\PB{\ppi_a(x_1)}{q_{bc}(x_2)}
&=&- \Omega(x_1)(\overline{D}_d^1-\varphi_d(x_1)) P^{d}{}_{abc}(x_1,x_2)
+ (k_b(x_1)q_{ac}(x_1)+k_c(x_1)q_{ab}(x_1))\del{1}{2}.\nn
\eeq
Integrating these equations over $x_1$ and contracting with $B^a(x_1)$ gives the action (\ref{LBS},\ref{LBp}) of $p_B$ on $\tilde\pi_a$ and $q_{ab}$. We can proceed similarly for the action of $p_B$ on $\mu$.
This provides an independent check of the
complete spin-$1$ sector.

The only remaining bracket is the one of $\mu$ with itself, which can be computed using $m_\beta$,
\begin{align}
\{\mu(x_1), \mu(x_2)\}
= \frac12 \sigma_{ab}(x_1) P^{ab cd}(x_1,x_2)\sigma_{cd}(x_2),
\end{align} 
which is exactly \eqref{mm}.
This is the final step in the independent verification of the third stage brackets, the main result of this paper. 

It is useful to conclude by relating our construction more precisely to two closely connected approaches. The pioneering analysis of Reisenberger \cite{Reisenberger:2007ku,Reisenberger:2012zq,Reisenberger:2018xkn} derives the Poisson brackets of a complete set of free gravitational data on a double-null sheet. In that formulation the null Einstein equations have already been solved: the resulting data are unconstrained and parametrize solutions in the domain of dependence of the double-null initial surface. In this approach the spin-0 and spin spin-1 degrees of freedom do not appear in the bulk of the null surfaces. Only the spin-2 degrees of freedom are considered, in particular, Reisenberger obtained the non-local self-bracket of the conformal two-metric along the null generators, which is the closest precursor of the spin-two bracket encoded here covariantly by the bilocal kernel $P_{ab}{}^{cd}$. The two constructions therefore operate at different stages of the reduction. Here we retain the Raychaudhuri and Damour equations as dynamical constraints and construct instead the kinematical Poisson structure on which they act. This keeps simultaneously the spin-zero, spin-one, and spin-two momenta and makes their Hamiltonian actions manifest. Conversely, the double-null formulation of \cite{Reisenberger:2007ku,Reisenberger:2012zq,Reisenberger:2018xkn} includes the codimension-two data at the intersection of the two null sheets and treats the issues associated with caustics and generator crossings explicitly, while our present bulk analysis is restricted to a single null segment between caustics and postpones the treatment of its corner degrees of freedom.

As we already mentioned, a closely related symplectic analysis was developed by Adami et al. in \cite{Adami:2023wbe}. The relation between the two parametrizations can be made explicit using the bulk metric reviewed in Appendix \ref{Ae}. Our bulk Gaussian-null line element
\beq
\rd s^2=2e^\alpha \rd v( \rd r-F \rd v)+q_{AB}( \rd \theta^A-U^A  \rd v)(\rd \theta^B-U^B\rd v)
\eeq
coincides with Eq. (2.1) of \cite{Adami:2023wbe} upon identifying
\beq
\eta=e^\alpha,\qquad V=2e^\alpha F,\qquad U^A_{[50]}=-U^A,\qquad x^A=\theta^A .
\eeq
In particular, the null generator and transverse metric agree directly, while their choice of Gaussian-null coordinates corresponds to the integrable Ehresmann gauge $k=\rd v$, for which $\varphi_A=\varpi_{AB}=0$. In four dimensions, on the null surface, their $\Omega_{AB}$, $\Omega$, and $\gamma_{AB}$ correspond respectively to our $q_{AB}$, $\Omega$, and $\bar q_{AB}$; in the co-rotating sector $U^A=0$, their news $N_{AB}=\frac12\partial_v\gamma_{AB}$ is therefore related to our shear by $\sigma_{AB}=\Omega N_{AB}$.

This makes the relation between the two symplectic analyses particularly transparent. At the off-shell level the underlying null data are closely related, but the essential distinction is the stage at which the Einstein equations are imposed. The main analysis of \cite{Adami:2023wbe} proceeds to the null-boundary solution phase space by solving the Raychaudhuri, Damour, and shear-evolution equations, and its explicit radiative reduction is performed in the co-rotating frame $U^A=0$. This leads to their decomposition into boundary and bulk symplectic structures, the associated gravitational flux, and the Carrollian geometry of the bulk solution space. Our purpose is complementary: we retain the intrinsic null constraints off shell and keep the spin-one pair associated with the horizontal components of the null generator and their conjugate momenta. Imposing the Raychaudhuri and Damour constraints and subsequently choosing the co-rotating gauge therefore provides a direct route from the kinematical phase space constructed here toward the solution-space description of \cite{Adami:2023wbe}.

\section{Conclusions}\label{s6}

In this work, we have derived the full kinematical Poisson structure of general relativity on a null hypersurface. Starting from the covariant symplectic potential, we first clarified the role of the symmetries entering the description of a ruled Carrollian geometry. In particular, we showed that shifts of the Ehresmann connection -- also referred to as Carroll boosts -- are pure gauge: they leave the null Brown-York stress tensor and the intrinsic Einstein equations invariant, carry no charge, and lie in the kernel of the presymplectic form. Consequently, the Ehresmann connection can be fixed as a background structure without restricting the physical diffeomorphisms of the null hypersurface. Compensating diffeomorphisms with appropriate rescalings and shifts then leads naturally to the prime phase space on which the physical gravitational data possess a non-degenerate symplectic structure.

We subsequently performed the complete kinematical reduction of this phase space, including the spin-zero, spin-one, and spin-two sectors. The resulting brackets describe the area element and surface tension, the null generator and momentum aspect, and the unimodular metric and shear. They make manifest the canonical roles of the corresponding momenta: brackets of the spin-zero momentum involve the evolution and reparametrizations along the null generators of the metric, whereas the spin-one momentum involves horizontal evolution and spatial diffeomorphisms on the cuts. The non-locality of the spin-two sector is encoded in a bilocal propagator that transports the metric degrees of freedom along the null hypersurface and organizes all the non-local contributions to the brackets. We derived these results through a three-step Dirac reduction and independently reconstructed them from the Hamiltonian generators of the phase-space transformations. The agreement between the two derivations provides a non-trivial check of the complete bracket structure. Our derivation was performed assuming $\cN$ is $3$-dimensional, but it can be easily generalized to any $d$ by using $\mu=\kappa+\frac{d-1}{d}\theta$ and $\Omega\to \Omega^{2/d}$, such that the symplectic structure is unchanged, and the third-step reduced brackets are unmodified.

The most immediate next step is dynamical. The brackets obtained here define the arena in which the Raychaudhuri and Damour equations must now be imposed as constraints. It will be important to determine their complete constraint algebra, perform the corresponding reductions, and construct the resulting null observables. This should also clarify the relation between the canonical description developed here and the thermodynamic and fluid-gravity interpretations of null hypersurface dynamics. A complete treatment of finite null regions will furthermore require incorporating non-vanishing corner fluxes and the associated edge degrees of freedom. Extending the analysis beyond individual null segments and understanding how the canonical data behave near, or across, caustics are equally important global questions.

Finally, the present results provide the classical starting point for the quantization of the full null gravitational phase space. Previous work treated the spin-zero sector and the spin-two sector perturbatively; the inclusion of the spin-one momentum and of the Damour constraint now makes it possible to investigate the coupled system of null generators. A central question is whether the anomaly encountered in the quantization of the Raychaudhuri constraint persists, is modified, or becomes part of a larger anomalous constraint algebra once transverse dynamics is included. Developing the corresponding quantum commutators, constructing dressed null observables as initiated in \cite{Freidel:2025ous, Freidel:2026stu}, and pursuing an algebraic quantization of the complete phase space as initiated in \cite{Ciambelli:2026vxa} constitute natural directions opened by the present analysis.

\paragraph{Acknowledgments}

We thank Martina Adamo, Rodrigo Andrade e Silva, Felipe Diaz, Temple He, Josh Kirklin, Marc Klinger, Puttarak Jai-akson, Luis Lehner, Michael Imseis, Rob Myers, Giulio Neri, Javi Peraza, Michael Reisenberger, Antoine Rignon-Bret, Leo Sanhueza, Simone Speziale, Themis Zikopoulos, and Kathryn Zurek for discussions.
We are grateful to the Banff International Research Station (BIRS) for our stays there. The result of the focused research group in November 2022 was \cite{Ciambelli:2023mir}, while that of the focused research group in November 2023 is this manuscript. We thank BIRS for the vibrant environment and wonderful atmosphere. Research at Perimeter Institute is supported in part by the Government of Canada through the  Department of Innovation, Science and Economic Development Canada and by the Province of  Ontario through the Ministry of Colleges and Universities. L.C. is supported by the Simons Collaboration on Celestial Holography.

\appendix

\section{Derivation of Damour Equation}\label{AppA}

It is instructive to provide a derivation of the Damour equation in \eqref{Damour0}, to show the typical manipulations involved in this manuscript. Using \eqref{tau}, we get
\beqn
8\pi G q_a{}^bD_cT_b{}^c&=&8\pi G q_a{}^bD_c(\tau_b \ell^c+\tau_b{}^c)\\
&=&q_a{}^b D_\ell\pi_b-\theta q_a{}^b D_\ell k_b+(\theta+\kappa)\pi_a+8\pi G q_a{}^b D_c\tau_b{}^c,\label{Dam1}
\eeqn
where we used $\tau_a= \frac{1}{8\pi G}(\pi_a -\theta k_a)$, equation \eqref{dl}, which gives $D_c\ell^c=\theta+\kappa$, and the fact that $q_a{}^b k_b=0$. From \eqref{Dk1} and using $\ell^c \omega_c=\kappa$, we gather
\beq
D_\ell k_b=-\kappa k_b-\pi_b-\varphi_b \quad \Rightarrow \quad q_a{}^bD_\ell k_b=-(\pi_a+\varphi_a),
\eeq
and thus $\pi_a+\varphi_a$ is the combination appearing in the horizontal parallel transport.
This, together with
\beq
q_a{}^b D_\ell\pi_b=q_a{}^b \cL_\ell \pi_b-q_a{}^b \pi_c D_b \ell^c=q_a{}^b \cL_\ell \pi_b-\theta_a{}^c \pi_c= \cL_\ell \pi_a- \frac{\theta}{2}\pi_a - \sigma_a{}^c\ \pi_c,
\eeq
allow us to further process \eqref{Dam1} and get
\beqn
8\pi G q_a{}^bD_cT_b{}^c=q_a{}^b \cL_\ell \pi_b- \sigma_a{}^c\pi_c+\left(\kappa+ \frac32 \theta  \right)\pi_a+\theta\varphi_a +8\pi G q_a{}^b D_c\tau_b{}^c. \label{Dam2}
\eeqn
We wish to manipulate the last term such that we can extract the horizontal covariant derivative. This can be done using that $\tau_b{}^c$ is a horizontal tensor,
\beq
q_a{}^b D_c \tau_b{}^c=q_a{}^b D_c(\tau_b{}^d q_d{}^c)=\tau_a{}^d D_c q_d{}^c+q_a{}^b  q_d{}^c D_c\tau_b{}^d=\tau_a{}^d D_c q_d{}^c+\overline{D}_d\tau_a{}^d,
\eeq
where we introduced the horizontal covariant derivative defined by projecting all the indices with the projector $q_a{}^b$ (satisfying $q_a{}^bq_b{}^c=q_a{}^c$)
\beq
\overline{D}_d\tau_a{}^d= q_a{}^b  q_e{}^c q_d{}^e D_c\tau_b{}^d = q_a{}^b  q_d{}^c D_c\tau_b{}^d. \label{Hder}
\eeq
Note that, for a scalar field $\phi$, one has
\be \label{commutator}
[\cL_\ell, \overline{D}_a]\phi= \varphi_a \cL_\ell\phi.
\ee 

Finally, thanks to 
\beq\label{tauq}
\tau_a{}^d D_c q_d{}^c=-\tau_a{}^dD_c(k_d\ell^c)=\tau_a{}^d(\pi_d+\varphi_d),
\eeq
and using $\mu=\kappa+\frac{\theta}{2}$, equation \eqref{Dam2} becomes
\beqn
8\pi G q_a{}^bD_cT_b{}^c &=& q_a{}^b \cL_\ell \pi_b- \sigma_a{}^c\pi_c+(\mu+\theta) \pi_a+\theta\varphi_a \\
&&+8\pi G (\tau_a{}^d(\pi_d+\varphi_d)+\overline{D}_d\tau_a{}^d)\\
&=& q_a{}^b \cL_\ell \pi_b- \sigma_a{}^c\pi_c+(\mu+\theta) \pi_a+\theta\varphi_a\\
&&+ (\sigma_a{}^d-\mu q_a{}^d)(\pi_d+\varphi_d)+\overline{D}_d(\sigma_a{}^d-\mu q_a{}^d)\\
&=&q_a{}^b \cL_\ell \pi_b+\theta \pi_a+\theta\varphi_a+\varphi_d(\sigma_a{}^d-\mu q_a{}^d)+\overline{D}_d(\sigma_a{}^d-\mu q_a{}^d)\\
&=&(\cL_\ell+\theta)\pi_a+\theta\varphi_a+(\overline{D}_b+\varphi_b)(\sigma_a{}^b-\mu q_a{}^b),\label{DamA}
\eeqn
where we used $q_a{}^b \cL_\ell \pi_b =\cL_\ell \pi_a$. This is exactly the vacuum contribution to the Damour equation in \eqref{Damour0}.

\section{Symplectic Reduction: Step 1 and 2}\label{app}

\subsection{First Step Reduction}\label{sec:First-step}

We want to first introduce $k_a$ as a background structure. The first constraint is 
\beq
K(x_1)=\ell^a(x_1)k_a(x_1)-1.
\eeq
Assuming that $\delta k_a=0$, this satisfies
\beq
\PB{K(x_1)}{K(x_2)}_0=0.
\eeq

It acts on the symplectic fields as
\beq
\PB{K(x_1)}{\ttau^{ab}(x_2)}_0=0\qquad \PB{K(x_1)}{q_{ab}(x_2)}_0=0\qquad \PB{K(x_1)}{\ell^a(x_2)}_0=0,
\eeq
and\footnote{We recall that, as in the main body, we assumed $8\pi G=1$.}
\beq
\PB{K(x_1)}{\ttau_{a}(x_2)}_0=k_a(x_1)\del{1}{2}.
\eeq
This suggests the introduction of the second constraint
\beq
L(x_1)=\ell^a(x_1)\ttau_a(x_1)-\ta(x_1).
\eeq
At this stage, $\ta$ is just a repackaging of the phase space variables, since $\delta\ta$ is free. To make this a true phase space constraint, we must impose conditions on $\ta$. Given the stress tensor \eqref{Thydro}, we see that $\tilde\tau_\ell$ plays the role of the densitized energy of the system. Moreover, from \eqref{tau}, we have $\ta=-\Omega\theta$, and thus $\ta$ is solely a function of $\ell^a$ and $q_{ab}$, $\ta=\ta[(q_{ab},\ell^a)]$.
This implies
\beq
\PB{\ta(x_1)}{q_{ab}(x_2)}_0=0\qquad \PB{\ta(x_1)}{\ell^{a}(x_2)}_0=0,\qquad 
\PB{\ta(x_1)}{\ta(x_2)}_0=0.
\eeq

Then, we compute
\beq\label{kl}
\PB{K(x_1)}{L(x_2)}_0=\ell^a(x_1)k_a(x_1)\del{1}{2}=K(x_1)\del{1}{2}+\del{1}{2}
\eeq
and
\beq\label{LL}
\PB{L(x_1)}{L(x_2)}_0=-\ell^a(x_1)\PB{\ttau_a(x_1)}{\ta(x_2)}_0-\ell^a(x_2)\PB{\ta(x_1)}{\ttau_a(x_2)}_0.
\eeq
The bracket in eq.~\eqref{kl} contains a piece proportional to the constraints and a piece which is non-vanishing on the phase space constraint surface. Therefore, $K$ and $L$ form a system of second-class constraints, and we can safely impose $K(x_1)=0$ in the RHS of eq.~\eqref{kl}, and $L(x_1)=0$ in the RHS of eq.~\eqref{LL}. In particular, since $\ta$ is a function of $q_{ab}$ and $\ell^a$ only, 
\beq
\PB{L(x_1)}{L(x_2)}_0&=&-\ell^a(x_1)\PB{\ttau_a(x_1)}{\ta(x_2)}_0-\ell^a(x_2)\PB{\ta(x_1)}{\ttau_a(x_2)}_0\\
&=&-\PB{L(x_1)}{\ta(x_2)}_0-\PB{\ta(x_1)}{L(x_2)}_0\\
&=&0.
\eeq

Putting $K$ first and $L$ second, the constraints matrix is thus
\beq
\Upsilon_{ij}(x_1,x_2)=\begin{pmatrix}
    0 & \del{1}{2}\\
    -\del{1}{2} & 0
\end{pmatrix},
\eeq
whose inverse is simply
\beq
(\Upsilon^{-1})^{ij}(x_1,x_2)=\begin{pmatrix}
    0 & -\del{1}{2}\\
    \del{1}{2} & 0
\end{pmatrix}.
\eeq
Note that $K$ commutes with all the fields except $\ttau_a$, while for $L$ we have
\beq
\PB{L(x_1)}{\ttau^{ab}(x_2)}_0&=&-\PB{\ta(x_1)}{\ttau^{ab}(x_2)}_0\\ \PB{L(x_1)}{q_{ab}(x_2)}_0&=&0\\
\PB{L(x_1)}{\ell^{a}(x_2)}_0&=&-\ell^a(x_1)\del{1}{2}\\
\PB{L(x_1)}{\ttau_{a}(x_2)}_0&=&\ttau_a(x_1)\del{1}{2}-\PB{\ta(x_1)}{\ttau_{a}(x_2)}_0.
\eeq

We have gathered all the information to compute the constrained brackets
\beq
\PB{A(x_1)}{B(x_4)}_1=\PB{A(x_1)}{B(x_4)}_0-\int_2\int_3 \PB{A(x_1)}{H_i(x_2)}_0 (\Upsilon^{-1})^{ij}(x_2,x_3)\PB{H_j(x_3)}{B(x_4)}_0,
\eeq
where $H_i=(K,L)$ collectively denotes the constraints, and Einstein implicit summation is assumed on the indices $i,j$, see \eqref{Dirac}. Moreover, we denoted with a subscript $1$ the constrained brackets, on which $H_i$ is implemented. This bracket is such that $\PB{L(x_1)}{\cdot}_1=0= \PB{K(x_1)}{\cdot}_1$, i.e., we can  impose the constraints strongly.
The final brackets after the first step reduction can be conveniently written as 
\beq
\PB{\ttau_a(x_1)}{\ttau_{b}(x_2)}_1&=&\big(\ttau_a(x_1)k_b(x_1)-k_a(x_1)\ttau_b(x_1)\big)\del{1}{2}\\
&&+k_a(x_1)\PB{\ta(x_1)}{\ttau_{b}(x_2)}_0+k_b(x_2)\PB{\ttau_{a}(x_1)}{\ta(x_2)}_0\nn\\
\PB{\ttau_a(x_1)}{\ttau^{bc}(x_2)}_1&=&k_a(x_1)\PB{\ta(x_1)}{\ttau^{bc}(x_2)}_0\\
\PB{\ttau_a(x_1)}{\ell^b(x_2)}_1&=&-q_a{}^b(x_1)\del{1}{2}\\
\PB{\ttau_a(x_1)}{q_{bc}(x_2)}_1&=&0\\
\PB{\ttau^{ab}(x_1)}{\ttau^{cd}(x_2)}_1&=&0\\
\PB{\ttau^{ab}(x_1)}{\ell^c(x_2)}_1&=&0\\
\PB{\ttau^{ab}(x_1)}{q_{cd}(x_2)}_1&=&2\delta^{ab}_{(cd)}\del{1}{2}\\
\PB{\ell^a(x_1)}{\ell^b(x_2)}_1&=&0\\
\PB{\ell^a(x_1)}{q_{bc}(x_2)}_1&=&0\\
\PB{q_{ab}(x_1)}{q_{cd}(x_2)}_1&=&0
\eeq
Given that the constraint $L(x)=0$ fixes the $\ell$-component of $\ttau_a$, we can  write 
\beq
\ttau_a=\ta k_a+\tilde\pi_a\qquad \ell^a\tilde\pi_a=0.
\eeq
Then, by systematically projecting these brackets to the space along and orthogonal to $\ell^a$,\footnote{Note the useful identities $\PB{q_{a}{}^{b}(x_1)}{q_{cd}(x_2)}_1=0=\PB{q_{a}{}^{b}(x_1)}{\ell^c(x_2)}_1$.} we deduce that the non-vanishing brackets among the fields $[q_{ab},\ell^a,\ttau^{ab},\tilde\pi_a]$ are given by
\beq
\PB{\ppi_a(x_1)}{\ppi_{b}(x_2)}_1&=&\big(\ppi_a(x_1)k_b(x_1)-k_a(x_1)\ppi_b(x_1)\big)\del{1}{2}\cr
\PB{\ppi_a(x_1)}{\ell^b(x_2)}_1&=&-q_a{}^b(x_1)\del{1}{2}\cr
\PB{\ttau^{ab}(x_1)}{q_{cd}(x_2)}_1&=&2\delta^{ab}_{(cd)}\del{1}{2}.
\eeq 
One can directly check that this bracket satisfies the Jacobi identity and is consistent with the constraint. Brackets involving the densitized energy can be computed from the fact that it is a composite field of $q_{ab}$ and $\ell^a$.

\subsection{Second Step Reduction}
\label{sec:Second-step}

We wish now to impose 2 more constraints $Q_b= T^a=0$ with
\beq
Q_b(x_1)=\ell^a(x_1)q_{ab}(x_1)\qquad T^a(x_1)=\ttau^{ab}(x_1)k_b(x_1).
\eeq
They express the orthogonality of $q_{ab}$ and $\ttau^{ab}$ with respect to the temporal direction. 

Using the first-step-reduced brackets  we find that these constraints form a second class system, with brackets 
\beq
&\PB{Q_a(x_1)}{Q_b(x_2)}_1=0\qquad 
\PB{T^a(x_1)}{T^b(x_2)}_1=0,&\\
&
\PB{Q_a(x_1)}{T^b(x_2)}_1=
-(q_a{}^b(x_1)+2\ell^b(x_1)k_a(x_1))\del{1}{2}.&
\eeq
Ordering the constraints as $S_i=(Q_a,T^b)$, the constraints matrix is 
\beq
 \Upsilon_{ij}(x_1,x_2)=\begin{pmatrix}
     0 & -(q_a{}^b(x_1)+2\ell^b(x_1)k_a(x_1))\del{1}{2}\\(q_a{}^b(x_1)+2\ell^b(x_1)k_a(x_1))\del{1}{2}  & 0
\end{pmatrix}.
\eeq
Its inverse is
\beq
 (\Upsilon^{-1})^{ij}(x_1,x_2)=\begin{pmatrix}
     0 & (q_a{}^b(x_1)+\frac12\ell^b(x_1)k_a(x_1))\del{1}{2}\\-(q_a{}^b(x_1)+\frac12\ell^b(x_1)k_a(x_1))\del{1}{2}  & 0
\end{pmatrix},
\eeq
which indeed satisfies
\beq
 \int_{2} \Upsilon_{ij}(x_1,x_2)(\Upsilon^{-1})^{jk}(x_2,x_3)=\delta^k_i\tilde\delta^{(3)}(x_1,x_3).
\eeq

We can now evaluate how the constraints act on the phase space variables:
\beq
\PB{Q_a(x_1)}{\ell^b(x_2)}_1&=&0\\
\PB{Q_a(x_1)}{q_{bc}(x_2)}_1&=&0\\
\PB{Q_a(x_1)}{\ppi_b(x_2)}_1&=& q_{ab}(x_1)\del{1}{2}\\
\PB{Q_a(x_1)}{\ttau^{bc}(x_2)}_1&=&
-(\delta_a^b\ell^c(x_1)+\delta_a^c\ell^b(x_1))\del{1}{2}\\
\PB{T^a(x_1)}{\ell^b(x_2)}_1&=&0\\
\PB{T^a(x_1)}{q_{bc}(x_2)}_1&=&
(k_b(x_1)\delta^a_c+k_c(x_1)\delta^a_b)\del{1}{2}\\
\PB{T^a(x_1)}{\ppi_b(x_2)}_1&=&0\\
\PB{T^a(x_1)}{\ttau^{bc}(x_2)}_1&=&0.
\eeq
With these, we can finally evaluate the second-step-reduced brackets. 
Since $\ell^a$ commutes with both constraints, the brackets that involve it are unaffected. Moreover, since $q_{ab}$ commutes with $Q_a$, its own bracket is not modified. A similar feature occurs for $\ttau^{ab}$ and $\ppi_a$, which commute with $T^a$, and thus the brackets of $(\ttau^{ab},\ppi_a)$ are unaffected and still vanish. Therefore, so far we have learned that the following brackets are unchanged:
\beq
\PB{\ell^a(x_1)}{\ell^b(x_2)}_2&=&0\\
\PB{\ell^a(x_1)}{q_{bc}(x_2)}_2&=&0\\
\PB{\ppi_a(x_1)}{\ell^b(x_2)}_2&=&-q_a{}^b(x_1)\del{1}{2}\\
\PB{\ppi_a(x_1)}{\ppi_{b}(x_2)}_2&=&\big(\ppi_a(x_1)k_b(x_1)-k_a(x_1)\ppi_b(x_1)\big)\del{1}{2}\\
\PB{\ppi_a(x_1)}{\ttau^{bc}(x_2)}_2&=& 0\\
\PB{\ttau^{ab}(x_1)}{\ell^c(x_2)}_2&=&0\\
\PB{q_{ab}(x_1)}{q_{cd}(x_2)}_2&=&0\\
\PB{\ttau^{ab}(x_1)}{\ttau^{cd}(x_2)}_2&=&0.
\eeq

The remaining brackets to evaluate are 
\beq
\PB{\ppi_a(x_1)}{q_{bc}(x_2)}_2\qquad 
\PB{\ttau^{ab}(x_1)}{q_{cd}(x_2)}_2,
\eeq
which are modified by the second-step constraints. Using \eqref{Dirac} (with $C=S$), after a long yet straightforward computation we find
\beq
\PB{\ppi_a(x_1)}{q_{bc}(x_4)}_2
=(k_b(x_1)q_{ac}(x_1)+k_c(x_1)q_{ab}(x_1))\del{1}{4},
\eeq
and
\beq
\PB{\ttau^{ab}(x_1)}{q_{cd}(x_4)}_2=(q_c{}^a(x_1)q_d{}^b(x_1)+q_d{}^a(x_1)q_c{}^b(x_1))\del{1}{4}.
\eeq
This completes our second-step reduction.

\section{Third Step Reduction}\label{AC}

This appendix details computations in support of the third step reduction performed in section \ref{sec42}.

\subsection{Inverse of the Constraint Bracket}\label{AC1}

We want to invert \eqref{DDD}. The inverse is encoded in the tensor $P_{abcd}(\tilde{x}_2,\tilde{x}_3):= \Omega(x_2)P_{abcd}(x_2,x_3)\Omega(x_3)$, which by construction must satisfy
\beq\label{Pinverse}
2 \del{ 1}{3}\Pi_{ab}{}^{ef}(x_1)&=&\int_2 \Upsilon_{abcd}(x_1,x_2)P^{cdef}(\tilde{x}_2,\tilde{x}_3)\\
&=&\int_2 \Big[\Big(\hat\tau_{ab}(x_1)\bq_{cd}(x_1)-\bq_{ab}(x_1)\hat\tau_{cd}(x_1)\Big)\del{ 1}{2}
\\&&+\Omega(x_1)\cL_{\ell(x_1)}(\overline{\Pi}_{abcd}(x_1)\del{ 1}{2})\\
&&-\Omega(x_2)\cL_{\ell(x_2)}(\overline{\Pi}_{abcd}(x_1)\del{ 1}{2})\Big]P^{cdef}(\tilde{x}_2,\tilde{x}_3)\\
&=&\Big(\hat\tau_{ab}(x_1)\bq_{cd}(x_1)-\bq_{ab}(x_1)\hat\tau_{cd}(x_1)\Big)P^{cdef}(\tilde{x}_1,\tilde{x}_3)\\
&&+\Omega(x_1) \cL_{\ell(x_1)}\left[\overline{\Pi}_{abcd}(x_1)\int_2 \del{ 1}{2}P^{cdef}(\tilde{x}_2,\tilde{x}_3)\right]\\
&&-\int_2\cL_{\ell(x_2)}\left[ \Omega(x_2) \overline{\Pi}_{abcd}(x_1)\del{ 1}{2}P^{cdef}(\tilde{x}_2,\tilde{x}_3)\right]\\
&&+\int_2 \overline{\Pi}_{abcd}(x_1)\del{ 1}{2}\cL_{\ell(x_2)}\left[ \Omega(x_2)P^{cdef}(\tilde{x}_2,\tilde{x}_3)\right]\\
&=&\Big(\hat\tau_{ab}(x_1)\bq_{cd}(x_1)-\bq_{ab}(x_1)\hat\tau_{cd}(x_1)\Big)P^{cdef}(\tilde{x}_1,\tilde{x}_3)\\
&&+\Omega(x_1)\cL_{\ell(x_1)}\left[\overline{\Pi}_{abcd}(x_1)P^{cdef}(\tilde{x}_1,\tilde{x}_3)\right]\\
&&+\overline{\Pi}_{abcd}(x_1)\cL_{\ell(x_1)}\left[ \Omega(x_1)P^{cdef}(\tilde{x}_1,\tilde{x}_3)\right]
\eeq
This is a differential equation for $P_{abcd}$. 
Since $\Omega^2(x_1)\overline{\Pi}_{abcd}(x_1)=\Pi_{abcd}(x_1)$ lowers the indices of tensors maintaining the same weight, one has
\beq
\overline{\Pi}_{abcd}(x_1)P^{cdef}(\tilde{x}_1,\tilde{x}_2)&=&\Omega^{-2}(x_1)P_{ab}{}^{ef}(\tilde{x}_1,\tilde{x}_2)
\eeq
and thus, after 
dividing everything by $\Omega(x_3)$ and imposing the constraint $\hat\tau_{ab}=\sigma_{ab}$, we get  
\beq
2 \delc{1}{3}\Pi_{ab}{}^{ef}(x_1)
&=&- q_{ab}(x_1)\sigma^{cd}(x_1)P_{cd}{}^{ef}(x_1,x_3)+\Omega(x_1)\cL_{\ell(x_1)}\left[\frac{P_{ab}{}^{ef}(x_1,x_3)}{\Omega(x_1)}\right]\\
&&+\Omega^{-2}(x_1)\Pi_{abcd}(x_1)\cL_{\ell(x_1)}\left[ \Omega^2(x_1)P^{cdef}(x_1,x_3)\right]\\
&=&-q_{ab}(x_1)\sigma^{cd}(x_1)P_{cd}{}^{ef}(x_1,x_3)+(2\cL_{\ell(x_1)}-\theta(x_1))P_{ab}{}^{ef}(x_1,x_3)\\
&&
-\Omega^{2}(x_1) \cL_{\ell(x_1)}\overline\Pi_{abcd}(x_1)P^{cdef}(x_1,x_3)
\eeq

This can be simplified further, observing that 
\beq
\Omega^2\cL_{\ell}\overline{\Pi}_{abcd}&=&\frac{\Omega^2}{2}
\cL_{\ell}\left(\bq_{ac}\bq_{bd}+\bq_{ad}\bq_{bc}-\bq_{ab}\bq_{cd}\right)\\
 &=&\sigma_{ac} q_{bd}+ q_{ac}\sigma_{bd}+\sigma_{ad} q_{bc}+ q_{ad}\sigma_{bc}-\sigma_{ab} q_{cd}- q_{ab}\sigma_{cd},
\eeq
such that
\beq
\delc{1}{3}\Pi_{ab}{}^{ef}(x_1)&=&-\frac12 q_{ab}(x_1)\sigma^{cd}(x_1)P_{cd}{}^{ef}(x_1,x_3)+(\cL_{\ell(x_1)}-\frac12\theta(x_1))P_{ab}{}^{ef}(x_1,x_3)\\
&&
-\frac12(2\sigma_{ac}(x_1) q_{bd}(x_1)+2 q_{ac}(x_1)\sigma_{bd}(x_1)- q_{ab}(x_1)\sigma_{cd}(x_1))P^{cdef}(x_1,x_3)\nn\\
&=&
(\cL_{\ell(x_1)}-\frac12 \theta(x_1))P_{ab}{}^{ef}(x_1,x_3)\\
&&-(\sigma_{ac}(x_1) q_{bd}(x_1)+ q_{ac}(x_1)\sigma_{bd}(x_1))P^{cdef}(x_1,x_3)
\eeq
Therefore, the final differential equation that the inverse of the constraint bracket must satisfy is
\beq
\delc{1}{3}\Pi_{ab}{}^{ef}(x_1)&=&\Big(\cL_{\ell(x_1)}-\frac{\theta(x_1)}{2}\Big)P_{ab}{}^{ef}(x_1,x_3)\\
&&-(\sigma_{ac}(x_1) q_{bd}(x_1)+ q_{ac}(x_1)\sigma_{bd}(x_1))P^{cdef}(x_1,x_3)\,.
\eeq
It can then be checked that both sides of this identity vanish after contracting with $ q^{ab}(x_1)$.

\subsection{Brackets with Constraint}\label{AC2}

In this appendix, we evaluate the bracket $\PB{\ppi_{a}(x_1)}{\chi_{cd}(x_2)}_2$. First, we have
\beq\label{pxh}
\PB{\ppi_{a}(x_1)}{\chi_{cd}(x_2)}_2=\PB{\ppi_{a}(x_1)}{\hat\tau_{cd}(x_2)}_2-\frac{\Omega(x_2)}{2}\PB{\ppi_{a}(x_1)}{\cL_{\ell(x_2)}\bq_{cd}(x_2)}_2\,.
\eeq
Therefore, we need the brackets
\beq
\PB{\ppi_a(x_1)}{\ell^b(x_2)}_2&=&-q_a{}^b(x_1)\del{1}{2}\\
\PB{\ppi_a(x_1)}{\bq_{bc}(x_2)}_2&=&(k_b(x_1)\bq_{ac}(x_1)+k_c(x_1)\bq_{ab}(x_1))\del{1}{2}.
\eeq

To evaluate $\PB{\ppi_{a}(x_1)}{\Omega(x_2)\cL_{\ell(x_2)}\bq_{cd}(x_2)}_2$, we use
\beq
\Omega\cL_{\ell}\bq_{bc}=\cL_{\ell}q_{bc}-\frac{q_{bc}}{2}q^{de}\cL_{\ell}q_{de}\,.
\eeq
To lighten the notation in these long computations, if the point specification is omitted it is intended to be
evaluated at $x_2$. Using the definition of the Lie derivative, one has
\beq
\PB{\ppi_a(x_1)}{\cL_{\ell}q_{bc}}&=&\del{1}{2}\left(-q_b{}^dq_c{}^e \pa_a q_{de}+k_a\cL_{\ell}q_{bc}+k_b\cL_{\ell}q_{ac}+k_c\cL_{\ell}q_{ab}\right)\\
&&-q_{ab}(\overline\pa_c +\varphi_c)\del{1}{2}-q_{ac}(\overline\pa_b +\varphi_b)\del{1}{2}\,.
\eeq
From this we evaluate
\beq
\PB{\ppi_a(x_1)}{q^{bc}\cL_{\ell}q_{bc}}&=&\del{1}{2}\left(-q^{bc}\pa_a q_{bc}+k_a q^{bc}\cL_{\ell}q_{bc}\right)-2(\overline \pa_a+\varphi_a)\del{1}{2}\,.
\eeq
Combining these two results we get
\beq
\PB{\ppi_a(x_1)}{\Omega\cL_{\ell}\bq_{bc}}
&=&\Omega \del{1}{2}\left(k_a\cL_{\ell}\bq_{bc}+k_b\cL_{\ell}\bq_{ac}+k_c\cL_{\ell}\bq_{ab}-q_b{}^dq_c{}^e \pa_a \bq_{de}\right)\\
&&+q_{bc}(\overline \pa_a+\varphi_a)\del{1}{2}-q_{ab}(\overline\pa_c +\varphi_c)\del{1}{2}-q_{ac}(\overline\pa_b +\varphi_b)\del{1}{2}\nn\\
&=&2\del{1}{2}\left(k_a\sigma_{bc}+k_b\sigma_{ac}+k_c\sigma_{ab}-\frac{\Omega}{2} q_b{}^dq_c{}^e \pa_a \bq_{de}\right)\nn\\
&&+q_{bc}(\overline \pa_a+\varphi_a)\del{1}{2}-q_{ab}(\overline\pa_c +\varphi_c)\del{1}{2}-q_{ac}(\overline\pa_b +\varphi_b)\del{1}{2}\,,\nn
\eeq
where we imposed the constraint in the last step.

Therefore, we have all the tools to evaluate \eqref{pxh}:
\beq
\PB{\ppi_a(x_1)}{\chi_{bc}}
&=&\del{1}{2}\big(k_a\sigma_{bc}-\frac{\Omega}{2} q_b{}^dq_c{}^e \pa_a \bq_{de}\big)\\
&&+\left(\frac{q_{bc}}{2}(\overline \pa_a+\varphi_a)-\frac{q_{ab}}{2}(\overline\pa_c +\varphi_c)-\frac{q_{ac}}{2}(\overline\pa_b +\varphi_b)\right)\del{1}{2}\,.
\eeq

\subsection{Final Step Reduction}\label{AC3}

The last reduction step is performed computing the Dirac brackets
\beq
\PB{A(x_1)}{B(x_4)}_3=\PB{A(x_1)}{B(x_4)}_2-\frac12 \int_2\int_3 \PB{A(x_1)}{\chi_{cd}(x_2)}_2 P^{cdef}(\tilde{x}_2,\tilde{x}_3)\PB{\chi_{ef}(x_3)}{B(x_4)}_2.
\eeq 
The factor $\frac12$ is there since we have chosen, in \eqref{Pinverse}, $\frac{P_{abcd}}{2}$ to be the inverse of the constraint bracket.

Since $\Omega$ and $\ell$ commute with the constraint, all the brackets involving them are unmodified:
\beq
\PB{\Omega(x_1)}{\Omega(x_2)}_3&=&0\\
\PB{\Omega(x_1)}{{\mu}(x_2)}_3&=&\del{1}{2}\\
\PB{\Omega(x_1)}{\ell^a(x_2)}_3&=&0\\
\PB{\Omega(x_1)}{\tilde\pi_{a}(x_2)}_3&=&0\\
\PB{\Omega(x_1)}{\overline q_{ab}(x_2)}_3&=&0\\
\PB{\ell^a(x_1)}{\mu(x_2)}_3&=&0\\
\PB{\ell^a(x_1)}{\ell^b(x_2)}_3&=&0\\
\PB{\ell^a(x_1)}{\ppi_b(x_2)}{}_3&=&q_b{}^a(x_1)\del{1}{2}\\
\PB{\ell^a(x_1)}{\bq_{bc}(x_2)}_3&=&0
\eeq

We start with the brackets involving $\mu$. Using the step-2 brackets, we evaluate
\beq
\PB{\pmu(x_1)}{\pmu(x_4)}_3&=&-\frac12 \int_2\int_3 \PB{\pmu(x_1)}{\chi_{cd}(x_2)}_2 P^{cdef}(\tilde{x}_2,\tilde{x}_3)\PB{\chi_{ef}(x_3)}{\pmu(x_4)}_2\\
&=& \frac12 \int_2\int_3 \delc{ 1}{2}\sigma_{cd}(x_2)P^{cdef}(\tilde{x}_2,\tilde{x}_3)\delc{3}{4} \sigma_{ef}(x_3)\\
&=&\frac12 \sigma_{cd}(x_1)P^{cdef}(x_1,x_4)\sigma_{ef}(x_4)\,,
\eeq
where we used $\del{1}{2}=\Omega(x_1)\delc{1}{2}$, \eqref{ddd} and $P_{abcd}(\tilde x_2,\tilde x_3)=\Omega(x_2)P_{abcd}(x_2,x_3)\Omega(x_3)
$.

Similarly (recall $\overline{\Pi}_{abcd}=\overline{\Pi}_{cdab}$)
\beq
\PB{\pmu(x_1)}{\bq_{ab}(x_4)}_3&=&-\frac12\int_2\int_3 \PB{\pmu(x_1)}{\chi_{cd}(x_2)}_2 P^{cdef}(\tilde{x}_2,\tilde{x}_3)\PB{\chi_{ef}(x_3)}{\bq_{ab}(x_4)}_2\\
&=&-\int_2\int_3 \sigma_{cd}(x_2) \delc{1}{2} P^{cdef}(\tilde{x}_2,\tilde{x}_3)\overline{\Pi}_{efab}(x_4)\del{3}{4}\\
&=&-\Omega(x_4)\sigma_{cd}(x_1)P^{cdef}(x_1,x_4) \overline{\Pi}_{efab}(x_4)\\
&=&-\frac{\sigma_{cd}(x_1) P^{cd}{}_{ab}(x_1,x_4)}{\Omega(x_4)}.
\eeq
Note that, using the commutator between $\mu$ and $\Omega$, we can write
\beq
\PB{\pmu(x_1)}{q_{ab}(x_4)}_3&=&\PB{\pmu(x_1)}{\Omega(x_4)}_3\bq_{ab}(x_4)+\Omega(x_4)\PB{\pmu(x_1)}{\bq_{ab}(x_4)}_3\\
&=&-q_{ab}(x_1)\delc{1}{4}-\sigma_{cd}(x_1) P^{cd}{}_{ab}(x_1,x_4)
\eeq

The last bracket to compute involving $\mu$ is (recall $P^{abcd}q_{cd}=0$)
\beq
\PB{\mu(x_1)}{\tilde\pi_{a}(x_4)}_3&=&-\frac12\int_2\int_3 \PB{\mu(x_1)}{\chi_{cd}(x_2)}_2 P^{cdef}(\tilde{x}_2,\tilde{x}_3)\PB{\chi_{ef}(x_3)}{\tilde\pi_a(x_4)}_2\\
&=&-\frac12\sigma_{cd}(x_1) P^{cdef}(x_1,x_4)\Omega(x_4)\left(\frac{1}{2}\pa_a^4 q_{ef}(x_4)-k_a(x_4)\sigma_{ef}(x_4)\right)\\
&&-\frac12\int_3\sigma_{cd}(x_1) P^{cde}{}_{a}(x_1,x_3)\Omega(x_3)(\overline\pa_e^3+\varphi_e(x_3))\del{3}{4}\\
&=&-\frac12\sigma_{cd}(x_1) P^{cdef}(x_1,x_4)\Omega(x_4)\left(\frac{1}{2}\pa_a^4 q_{ef}(x_4)-k_a(x_4)\sigma_{ef}(x_4)\right)\\
&&+\frac12\paleft_e^4\left(\sigma_{cd}(x_1) P^{cde}{}_{a}(x_1,x_4)\Omega(x_4)\right)\,,
\eeq
where, to lighten notation, we defined $\paleft^4_a=\pa^4_a-\varphi_a(x_4)$, which features after integration by parts.
Now one has that, once contracted with $P^{cdef}$,
\beq\label{lbq}
(\frac12 \pa_aq_{ef}-k_a \sigma_{ef})P^{cdef}=\frac12 (\pa_aq_{ef}-k_a\cL_{\ell}q_{ef})P^{cdef}=\frac12 \cL_{\overline\pa_a}q_{ef}P^{cdef}
\eeq
and thus
\beq
\PB{\mu(x_1)}{\tilde\pi_{a}(x_4)}_3&=&-\frac14\sigma_{cd}(x_1) P^{cdef}(x_1,x_4)\Omega(x_4)\cL_{\overline\pa^4_a}q_{ef}(x_4)\\
&&+\frac12\paleft_e^4\left(\sigma_{cd}(x_1) P^{cde}{}_{a}(x_1,x_4)\Omega(x_4)\right)\,.
\eeq

We then move to $\ppi_{a}$. 
There are two commutators to evaluate. The first one is $\PB{\tilde\pi_{a}(x_1)}{\tilde\pi_{b}(x_4)}_3$, which is also the longest computation in this appendix. We start noticing that, from the previous computation, we have learned
\beq
P^{cdef}(\tilde x_2,\tilde x_3)\PB{\chi_{ef}(x_3)}{\ppi_{h}(x_4)}_2&=&\frac12 P^{cdef}(\tilde x_2,\tilde x_3)\cL_{\overline\pa_h^3}q_{ef}(x_3)\del{3}{4}\\
&&+P^{cde}{}_{h}(\tilde x_2,\tilde x_3)(\pa_e^3+\varphi_e(x_3))\del{3}{4}\,,
\eeq
and, similarly,
\beq
\PB{\ppi_{a}(x_1)}{\chi_{cd}(x_2)}_2P^{cdef}(\tilde x_2,\tilde x_3)&=&-\frac12 P^{cdef}(\tilde x_2,\tilde x_3)\cL_{\overline\pa_a^2}q_{cd}(x_2)\del{1}{2}\\
&&-P_a{}^{def}(\tilde x_2,\tilde x_3)(\pa_d^2+\varphi_d(x_2))\del{1}{2}\,.
\eeq
Therefore, we have
\beq
\PB{\ppi_a(x_1)}{\ppi_{h}(x_4)}_3&=&(\ppi_a(x_1) k_h(x_1)-\ppi_h(x_1)k_a(x_1))\del{1}{4}\\
&&-\frac12\int_2\int_3 \PB{\ppi_{a}(x_1)}{\chi_{cd}(x_2)}_2 P^{cdef}(\tilde{x}_2,\tilde{x}_3)\PB{\chi_{ef}(x_3)}{\ppi_{h}(x_4)}_2\\
&=&(\ppi_a(x_1) k_h(x_1)-\ppi_h(x_1)k_a(x_1))\del{1}{4}\\
&&-\frac14\int_2\PB{\ppi_{a}(x_1)}{\chi_{cd}(x_2)}_2 P^{cdef}(\tilde x_2,\tilde x_4)\cL_{\overline\pa_h^4}q_{ef}(x_4)\\
&&-\frac12\int_2\PB{\ppi_{a}(x_1)}{\chi_{cd}(x_2)}_2\int_3  P^{cde}{}_{h}(\tilde x_2,\tilde x_3)(\pa_e^3+\varphi_e(x_3))\del{3}{4}\\&=&(\ppi_a(x_1) k_h(x_1)-\ppi_h(x_1)k_a(x_1))\del{1}{4}\\
&&-\frac14\int_2\PB{\ppi_{a}(x_1)}{\chi_{cd}(x_2)}_2 P^{cdef}(\tilde x_2,\tilde x_4)\cL_{\overline\pa_h^4}q_{ef}(x_4)\\
&&+\paleft_e^4\frac12\int_2\PB{\ppi_{a}(x_1)}{\chi_{cd}(x_2)}_2 P^{cde}{}_{h}(\tilde x_2,\tilde x_4)\\
&=&(\ppi_a(x_1) k_h(x_1)-\ppi_h(x_1)k_a(x_1))\del{1}{4}\\
&&+\frac18 \cL_{\overline\pa_a^1}q_{cd}(x_1)P^{cdef}(\tilde x_1,\tilde x_4)\cL_{\overline\pa_h^4}q_{ef}(x_4)\\
&&+\frac14\int_2 P_a{}^{def}(\tilde x_2,\tilde x_4)(\pa_d^2+\varphi_d(x_2))\del{1}{2}\cL_{\overline\pa_h^4}q_{ef}(x_4)\\
&&-\frac14\paleft_e^4P^{cde}{}_h(\tilde x_1,\tilde x_4)\cL_{\overline\pa_a^1}q_{cd}(x_1)\\
&&-\frac12\paleft_e^4\int_2\Big(P_a{}^{de}{}_h(\tilde x_2,\tilde x_4)(\pa_d^2+\varphi_d(x_2))\del{1}{2}\Big)\\
&=&(\ppi_a(x_1) k_h(x_1)-\ppi_h(x_1)k_a(x_1))\del{1}{4}\\
&&+\frac18 \cL_{\overline\pa_a^1}q_{cd}(x_1)P^{cdef}(\tilde x_1,\tilde x_4)\cL_{\overline\pa_h^4}q_{ef}(x_4)+\frac12\paleft_d^1 \paleft_e^4 P_a{}^{de}{}_h(\tilde x_1,\tilde x_4)\\
&&-\frac14\paleft_d^1 P_a{}^{def}(\tilde x_1,\tilde x_4)\cL_{\overline\pa_h^4}q_{ef}(x_4)-\frac14\paleft_e^4P^{cde}{}_h(\tilde x_1,\tilde x_4)\cL_{\overline\pa_a^1}q_{cd}(x_1)
\eeq

Using again \eqref{lbq}, the second commutator involving $\ppi_a$ can be processed into
\beq
\PB{\ppi_a(x_1)}{\bq_{gh}(x_4)}_3&=&(k_g(x_1)\bq_{ah}(x_1)+k_h(x_1)\bq_{ag}(x_1))\del{1}{4}\\
&&-\frac12 \int_2\int_3 \PB{\ppi_{a}(x_1)}{\chi_{cd}(x_2)}_2 P^{cdef}(\tilde{x}_2,\tilde{x}_3)\PB{\chi_{ef}(x_3)}{\bq_{gh}(x_4)}_2\\
&=&(k_g(x_1)\bq_{ah}(x_1)+k_h(x_1)\bq_{ag}(x_1))\del{1}{4}\\
&&-\int_2
\PB{\ppi_{a}(x_1)}{\chi_{cd}(x_2)}_2 \frac{\Omega(x_2)P^{cd}{}_{gh}(x_2,x_4)}{\Omega(x_4)}\\
&=&(k_g(x_1)\bq_{ah}(x_1)+k_h(x_1)\bq_{ag}(x_1))\del{1}{4}\\
&&-\big(k_a(x_1)\sigma_{cd}(x_1)-\frac{\Omega(x_1)}{2}q_c{}^e(x_1)q_d{}^f(x_1) \pa_a^1 \bq_{ef}(x_1)\big)\frac{\Omega(x_1)P^{cd}{}_{gh}(x_1,x_4)}{\Omega(x_4)}\nn\\
&&+\int_2 \frac{\Omega(x_2)P^{c}{}_{agh}(x_2,x_4)}{\Omega(x_4)}(\overline\pa^2_c +\varphi_c(x_2))\del{1}{2}\\
&=&(k_g(x_1)\bq_{ah}(x_1)+k_h(x_1)\bq_{ag}(x_1))\del{1}{4}\\
&&+\frac{\Omega(x_1)}{\Omega(x_4)}\left(\frac12 \pa_a^1 q_{cd}(x_1)-k_a(x_1)\sigma_{cd}(x_1)\right) P^{cd}{}_{gh}(x_1,x_4)\\
&&-\frac{1}{\Omega(x_4)}\paleft^1_c\left(\Omega(x_1) P^{c}{}_{agh}(x_1,x_4)\right)\\
&=&(k_g(x_1)\bq_{ah}(x_1)+k_h(x_1)\bq_{ag}(x_1))\del{1}{4}+\frac{\Omega(x_1)}{2\Omega(x_4)}\cL_{\overline\pa^1_a}q_{cd}(x_1) P^{cd}{}_{gh}(x_1,x_4)\nn\\
&&-\frac{1}{\Omega(x_4)}\paleft^1_c\left(\Omega(x_1) P^{c}{}_{agh}(x_1,x_4)\right)
\eeq
multiplying both sides by $\Omega(x_4)$ we get, after renaming some variables
\beq
\PB{\ppi_a(x_1)}{q_{bc}(x_2)}_3&=&(k_b(x_1)q_{ac}(x_1)+k_c(x_1)q_{ab}(x_1))\del{1}{2}+\frac{\Omega(x_1)}{2}\cL_{\overline\pa^1_a}q_{de}(x_1) P^{de}{}_{bc}(x_1,x_2)\nn\\
&&-\paleft^1_d\left(\Omega(x_1) P^{d}{}_{abc}(x_1,x_2)\right)
\eeq

The last field remaining is $\bq_{ab}$, and the only missing commutator is the bracket among itself. One has
\beq
\PB{\bq_{ab}(x_1)}{\bq_{gh}(x_4)}_3&=&-\frac12 \int_2\int_3 \PB{\bq_{ab}(x_1)}{\chi_{cd}(x_2)}_2 P^{cdef}(\tilde{x}_2,\tilde{x}_3)\PB{\chi_{ef}(x_3)}{\bq_{gh}(x_4)}_2\\
&=&\frac12 \int_2\int_3 2\overline{\Pi}_{abcd}(x_1)\del{ 1}{2} P^{cdef}(\tilde{x}_2,\tilde{x}_3)2\overline{\Pi}_{efgh}(x_4)\del{3}{4}\\
&=&2\overline{\Pi}_{abcd}(x_1) P^{cdef}(\tilde{x}_1,\tilde{x}_4)\overline{\Pi}_{efgh}(x_4)\\
&=&2\frac{P_{abgh}(x_1,x_4)}{\Omega(x_1)\Omega(x_4)}
\eeq

Note that, since $\bq_{ab}$ commutes with $\Omega$, we can bring the denominator to the LHS and get 
\beq
\PB{q_{ab}(x_1)}{q_{gh}(x_4)}_3=2 P_{abgh}(x_1,x_4)\,,
\eeq
which is the pleasant expression reported in the main body of the manuscript.

\section{Variational 
Identity}\label{sec:variation}
Here we give a detailed proof of the variational identity \eqref{varPi2}. Let $X_{ab}=\frac{X}{2}q_{ab}+\hat X_{ab}$ e an arbitrary horizontal symmetric tensor, then
\beq 
\delta \Pi^{abcd} X_{cd} &=&- \frac{1}{2}\left(2\Delta^{ac}q^{bd} + 2\Delta^{bd} q^{ac} - \Delta^{ab}q^{cd}-q^{ab} \Delta^{cd}\right) X_{cd}
\\
&=& - \left((\Delta\cdot X)^{ab} +  (\Delta\cdot X)^{ba} - \frac{1}{2}\Delta^{ab}X-\frac{1}{2} q^{ab} (\Delta\!:\!X)\right) \\
&=& - \left(\Delta^{ab}X +  \Delta \hat X^{ab} + (\hat\Delta\!:\!\hat X) q^{ab} - \frac{1}{2}\Delta^{ab}X-\frac{1}{2}q^{ab} (\Delta\!:\!X)\right) \\
&=& - \left(\frac12 \Delta^{ab}X +  \Delta \hat X^{ab} + \frac12(\hat\Delta\!:\!\hat X) q^{ab} -\frac{1}{4}q^{ab} \Delta X\right) \\
&=& - \left(\frac12 \hat{\Delta}^{ab}X +  \Delta \hat X^{ab} + \frac12(\hat\Delta\!:\!\hat X) q^{ab} \right) \\
&=& - \left(\frac12 \hat{\Delta}^{ab} q^{cd}  + \frac12 q^{ab}\hat{\Delta}^{cd} +\Delta \Pi^{abcd} \right)    X_{cd}.
\eeq
One uses that 
\be 
(\Delta \!\cdot\! X)^{(ab)} = \frac12\left(
\Delta^{ab}X +  \Delta \hat X^{ab} + (\hat\Delta\!:\!\hat X) q^{ab}\right).
\ee 
Since the derivation above holds for an arbitrary horizontal symmetric tensor, it establishes the tensorial identity
\be 
\delta \Pi^{abcd} = - \left(\frac12 \hat{\Delta}^{ab} q^{cd}  + \frac12 q^{ab}\hat{\Delta}^{cd} +\Delta \Pi^{abcd} \right).
\ee

\section{Bulk Metric Parametrization}\label{Ae}

For completeness, we review in this Appendix the explicit expressions of the coefficients $(\mu,\pi_a,\theta_{AB})$ in terms of the bulk metric (see e.g \cite{Freidel:2024emv}).

Let $(v,r,\theta^A)$ be local Gaussian normal coordinates and consider the spacetime metric 
\begin{equation}
 \rd s^2
 =2e^{\alpha}\,\rd v\bigl(\rd r-F\,\rd v\bigr)
 +q_{AB}\bigl(\rd\theta^A-U^A\rd v\bigr)
          \bigl(\rd\theta^B-U^B\rd v\bigr).
 \label{eq:metric}
\end{equation}
We assume that the null surface is located at $r=0$, and on the null surface we have that  $F|_{\mathcal{N}}=0$.
We chose  the frame
\begin{equation}
 \ell=\pa_v+U^A\pa_A+F\pa_r \stackrel{\mathcal{N}}{=}\pa_v+U^A\pa_A,
 \qquad
 \hat{k}=e^{-\alpha}\pa_r,
 \qquad
 e_A=\pa_A.
 \label{eq:frame}
\end{equation}
with associated coframe 
\begin{align}
 \underline{\ell}
   =e^\alpha\bigl(\rd r-F\rd v\bigr),\qquad 
 k= \rd v,\qquad 
 e^A
   = \rd\theta^B-U^B\rd v .
   \end{align}
This choice of frame is such that $\varphi_A=0$ and $\varpi_{AB}=0$.
We can then evaluate the explicit expressions using the Koszul formula for the connection coefficients and we get that the value of the coefficients on $\mathcal{N}$ are
\beq
 \kappa &=&\ell[\alpha]+\pa_rF,
 \\
 \theta
 &=&\ell[\ln\sqrt{q}]
          +\pa_AU^A,\\
 \pi_A
 &=&\frac12\left(\pa_A\alpha
       +e^{-\alpha}q_{AB}\pa_rU^B\right),
 \\
 \theta_{AB}
 &=&\frac12 \pa_vq_{AB}+D_{(A}U_{B)},
\eeq
while 
\be 
\mu
 =\ell\left[\alpha +\tfrac12 \ln(\sqrt{q})\right]+\pa_rF
  +\frac12 \pa_AU^A.
\ee

\bibliographystyle{uiuchept}
\bibliography{CFLv1Damour.bib}

\end{document}